\documentclass[usenatbib]{mnras}
\usepackage[english]{babel}

\pdfoutput=1
\usepackage{amsfonts}
\usepackage{amsmath}
\usepackage{amssymb}
\usepackage{xcolor}
\usepackage{enumitem}
\usepackage{graphicx}
\usepackage{gensymb}
\usepackage{fancyhdr}
\usepackage{times}
\usepackage{widetext}
\usepackage{ulem}
\usepackage{caption}
\usepackage{subcaption} 
\usepackage{float} 

\newcommand{\LCDM}{$\Lambda$CDM}
\newcommand{\new}[1]{{\color{black} #1}}

\graphicspath{{./Figures/}}

\begin{document}

\title{Enhancing cosmic shear with the multi-scale lensing PDF}

\author[Giblin et al.]
{Benjamin Giblin$^{1,2}$\thanks{bengib@roe.ac.uk}, Yan-Chuan Cai$^2$ \& Joachim Harnois-D{\'e}raps$^3$  \\
$^1$ Institut de Ci{\`e}ncies del Cosmos, Universitat de Barcelona, Mart{\'i} i Franqu{\`e}s 1, Barcelona E08028,
Spain \\
$^2$ Institute for Astronomy, University of Edinburgh, Blackford Hill, Edinburgh, EH9 3HJ, UK \\
$^3$ School of Mathematics, Statistics and Physics, Newcastle University, Herschel Building, NE1 7RU, Newcastle-upon-Tyne, UK}

\maketitle
\begin{abstract}
We quantify the cosmological constraining power of the  `lensing PDF' - the one-point probability density of weak lensing convergence maps - by modelling this statistic numerically with an emulator trained on $w$CDM cosmic shear simulations. After validating our methods on Gaussian and lognormal fields, we show that `multi-scale' PDFs - measured from maps with multiple levels of smoothing - offer considerable gains over two-point statistics, owing to their ability to extract non-Gaussian information: for a mock Stage-III survey, lensing PDFs yield 33\% tighter constraints on the clustering parameter $S_8=\sigma_8\sqrt{\Omega_{\rm m}/0.3}$ than the two-point shear correlation functions. For Stage-IV surveys, we achieve $>$90\% tighter constraints on $S_8$, but also on the Hubble and dark energy equation of state parameters. Interestingly, we find improvements when combining these two probes only in our Stage-III setup; in the Stage-IV scenario the lensing PDFs contain all information from the standard two-point statistics and more. This suggests that while these two probes are currently complementary, the lower noise levels of upcoming surveys will unleash the constraining power of the PDF.
\end{abstract} 


\section{Introduction}\label{sec:introduction}

The field of cosmology currently stands at a crossroads. On the one hand, we have the standard cosmological model, \LCDM. This theoretically-motivated framework, when used to interpret astronomical observations, leads to the conclusion that more than 95\% of the Universe's energy density is contained in materials which remain poorly understood: cold dark matter (CDM) and an expansion-accelerating component, $\Lambda$, the cosmological constant. {\LCDM} has been highly successful in describing an impressive range of large-scale structure observations, from galaxy clustering \citep{satpathy/etal:2017} to the temperature and polarisation fluctuations of the CMB \citep{planck/etal:2018}, and many others \citep[see, for example,][]{huterer/etal:2018}. 

On the other hand, recent studies have cast doubt on the standard model through discrepancies arising in the comparison of cosmological constraints from different probes. The statistical significance of the disagreements between early- and late-time estimates of the Hubble parameter, for example, currently stand at the level of 4-6$\sigma$ depending on the data sets considered \citep{verde/etal:2019, riess/etal:2020, di-valentino/etal:2021}. Further cosmological tension can be found in the estimates of the clustering amplitude given by $S_8=\sigma_8 \sqrt{ \Omega_{\rm m}/0.3 }$, where $\sigma_8$ is the standard deviation of density fluctuations measured in spheres, and $\Omega_{\rm m}$ is the matter energy density parameter. In recent years, numerous weak gravitational lensing surveys \citep{heymans/etal:2013, jee/etal:2016, hildebrandt/etal:2017, troxel/etal:2018, hamana/etal:2019} measuring the \textit{cosmic shear} of galaxies at relatively low redshift $(z \lesssim 1.2)$ have consistently inferred lower values of this parameter than those obtained via the analyses of the much higher redshift $(z\sim 1100)$ CMB \citep{planck/etal:2016, planck/etal:2018}. Indeed, the three most advanced cosmic shear experiments ever undertaken, the Hyper Suprime-Cam Collaboration (HSC), the Kilo-Degree Survey (KiDS), and the Dark Energy Survey (DES), have all recently arrived at $S_8$ estimates 2-3$\sigma$ lower than the CMB-inferred values \citep{hikage/etal:2019, asgari/etal:2021, amon/etal:2021, secco/etal:2021}.   

The differences in cosmological parameter estimates between high- and low-redshift probes have prompted investigations into extensions to the standard model \citep{verde/etal:2013, joudaki/etal:2016, mortsell/dhawan:2018, des/etal:2018, troester/etal:2021}. However, no study to date has succeeded in identifying new physics which can simultaneously resolve the tensions seen under the fiducial {\LCDM} model whilst being consistent with observational constraints. Of course, there are other explanations for the observed cosmological tensions, including the simple possibility of unaccounted-for systematics in the analyses, or that the discrepancies are purely the result of a statistical fluctuation in the Universe's cosmic evolution. Whatever the truth may be, improving the precision of cosmological constraints will only deepen our insights into the nature of the cosmological discordance. With that in mind, this body of work is devoted to enhancing the precision of cosmic shear studies specifically, {by going beyond the conventional analysis based on two-point statistics, given that they are known to be sub-optimal in the non-Gaussian regime.}

The observables used to constrain parameters within a cosmic shear framework are the shapes of galaxies featuring coherent, correlated distortions, or \textit{shear}. The shear arises from the weak gravitational lensing of the galactic light by the foreground large-scale structure, and is only $\sim 1\%$ the amplitude of the intrinsic ellipticity of a typical galaxy \citep{bartelmann/schneider:2001}. Hence the lensing signal is only detectable statistically using very large samples of galaxies; concurrent Stage-III surveys analyse $\gtrsim 10^7$ sources \citep{giblin/etal:2021, gatti/etal:2021b}. Conventionally, the shear correlation function, which quantifies the strength of galaxy shape correlations as a function of angular separation and redshift, or its Fourier transform, the lensing power spectrum, are used to constrain cosmological parameters, with lensing showing the greatest sensitivity to $S_8$. However, two-point statistics such as these fail to extract all cosmological information from the data due to the non-Gaussian distribution of projected matter density on small scales, which arises from density perturbations in the late-time Universe. Given that our current observational data probes the non-Gaussian small-scale regime, the precision of our cosmological constraints is bottle-necked by the use of two-point statistics and its modelling. To further improve precision with current data sets, cosmic shear practitioners must venture beyond these standard probes.

Indeed, numerous studies have demonstrated the power and feasibility of cosmic shear with less conventional statistics. These include the use of three- and four-point functions \citep{takada/jain:2002, kilbinger/schneider:2005, fu/etal:2014, gatti/etal:2021, heydenreich/etal:2022}, the abundance and/or clustering of weak lensing overdensities \citep{jain/van-waerbeke:2000, kratochvil/etal:2010, liu/etal:2015, shan/etal:2018, martinet/etal:2018, harnois-deraps/etal:2021, davies/etal:2021} and underdensities \citep{davies/etal:2020}, density split statistics \citep{gruen/etal:2018, friedrich/etal:2018, brouwer/etal:2018, burger/etal:2022}, clipping the lensing signal from high-density regions \citep{giblin/etal:2018}, 
\new{Minkowski functionals \citep{kratochvil/etal:2012, petri/etal:2013, petri/etal:2015, grewal/etal:2022} 
and higher order moments \citep{wang/etal:2009, petri/etal:2013, petri/etal:2015, petri/etal:2016, vicinanza/etal:2016, heydenreich/etal:2021}} of weak lensing fields, 
and employing convolutional neural networks \citep{fluri/etal:2019}. Many of these ``beyond two-point" statistics have been shown to outperform the standard two-point functions as a direct consequence of the extraction of additional non-Gaussian cosmological information. 

The appeal of these new methods is clear, but their implementation does not come without challenges. First of all, many beyond two-point statistics do not have an analytical prescription, and those that do \citep[three-point, four-point, and density split statistics for example,][]{takada/jain:2002, fu/etal:2014, gruen/etal:2018} are not analytically modelled on small scales as accurately as the conventional two-point probes, limiting their potential cosmological precision. Generally this necessitates that numerical simulations are used to model the scale and cosmological dependence of beyond two-point statistics (although it is worth noting that two-point functions generally also require simulations to calibrate the small-scale signal). Secondly, the impact on beyond two-point statistics from various lensing systematics, including intrinsic galaxy alignments \citep{joachimi/bridle:2010, heymans/etal:2013}, baryonic feedback \citep{semboloni/etal:2011b}, survey masking \citep{giblin/etal:2018}, galaxy redshift errors \citep{asgari/etal:2021, secco/etal:2021}, and galaxy shape noise, generally needs to be modelled more thoroughly before they could replace two-point statistics as the default analysis tools in cosmic shear studies. 

In spite of these challenges, we explore the cosmological constraining power of the one-point probability density function (PDF) of the projected matter density (convergence) inferred through weak lensing. 
\new{The ``lensing PDF" is closely related to some of the aforementioned beyond two-point statistics; the first Minkowski functional, for example, is equivalent to the cumulative PDF, whereas the $n$'th moment of the convergence field is directly probed by the PDF sampled in $n$ convergence bins.} 

The cosmological dependence of the lensing PDF has recently been modelled analytically with large-deviation theory, and found to be consistent with the results from numerical simulations at the percent-level on mildly non-linear scales \citep[angular scales $>10$ arcminutes;][]{boyle/etal:2021}. The covariance of this statistic has also been accurately modelled following the same approach in \citet{uhlemann/etal:2022}.
The accuracy of this method relies on a transformation developed in \cite{bernardeau/etal:2014}, which nullifies the signal contribution of sources at a given redshift, allowing one to restrict the redshift distributions in cosmic shear observations to a definite interval. This in turn facilitates for the construction of lensing maps with limited contributions from highly non-linear scales. An alternative method of modelling the lensing PDF and its covariance using a halo-model formalism can also be found in \cite{thiele/etal:2020}, albeit with worse accuracy on the mildly non-linear scales. These studies find, through the use of Fisher matrix analyses, the cosmological constraining power of the lensing PDF to be comparable with the standard two-point shear correlation functions.

In contrast to these prior investigations, we model the lensing PDF cosmological dependence and covariance using only suites of numerical simulations\new{, an approach which was also adopted by \citet{liu/madhavacheril:2019}}. For the cosmological dependence we use the cosmo-SLICS \citep{harnois-deraps/etal:2019} which feature 26 distinct input cosmologies, and for the covariance, SLICS \citep{harnois-deraps/etal:2018}, consisting of more than 600 independent realisations of a single cosmology. These allow for an accurate measurement of the lensing PDF down to highly non-linear scales - the regime in which this statistic is expected to extract non-Gaussian information beyond the reach of conventional probes. 
Following \citet{boyle/etal:2021} \new{and \citep{kratochvil/etal:2012}}, we measure the lensing PDFs from projected density fields with various levels of smoothing and combine these statistics in our likelihood evaluation, in order to extract information on a range of physical scales and optimise the PDFs' constraining power.

We use a Gaussian process regression emulator trained on our simulated measurements to perform a full MCMC likelihood analysis. This leads to a more robust determination of the lensing PDFs' cosmological constraining power than is possible with Fisher matrix analyses. Furthermore, we perform these tests with simulations tailored to match the specifications of a concurrent Stage-III cosmic shear survey \citep{giblin/etal:2021, hildebrandt/etal:2021}, and with simulations of an upcoming stage-IV survey so as to demonstrate both the present-day feasibility and the future potential power of the lensing PDF. In doing so, we apply realistic galaxy shape noise and redshift binning, measuring both the ``auto" and ``cross" PDFs within each bin and from the combinations of bins respectively, thereby simulating a fully tomographic cosmic shear analysis with this new statistic.

\new{Our approach to modelling the lensing PDFs -- using an emulator trained on numerical simulations, as well as the implementation of tomography with a stage-IV mock survey -- bears similarities to \citet{liu/madhavacheril:2019}. There are also important differences in the two methodologies, including our use of the PDFs with multiple smoothing scales, our comparison of the current and future prospects of this statistic using different-generation mock surveys, and some dissimilarities in our simulations, elucidated in the following section.}

This paper is structured as follows. In Section \ref{sec:sims} we present the specifications of the numerical simulations we employ; in Section \ref{sec:method} we outline our methodology for measuring and modelling the lensing PDF, including validation tests of our approach with mock Gaussian and lognormal density fields; we present our results in Section \ref{sec:results}, and we conclude in Section \ref{sec:conclusions}.


\section{Simulations} \label{sec:sims}

The cosmological simulation suites known as SLICS and cosmo-SLICS \citep{harnois-deraps/etal:2018, harnois-deraps/etal:2019} have been employed extensively to estimate the cosmological dependence and covariance of novel weak lensing statistics \citep{martinet/etal:2018, giblin/etal:2018, davies/etal:2020, davies/etal:2021, martinet/etal:2021, harnois-deraps/etal:2021, heydenreich/etal:2021, burger/etal:2022}. Their efficacy in this regard accounts from the considerable number of simulation realisations, the relatively large simulation box size, and the tailoring of these mocks to present and future cosmic shear surveys.

\begin{figure}
\centering
\includegraphics[width=0.49\textwidth]{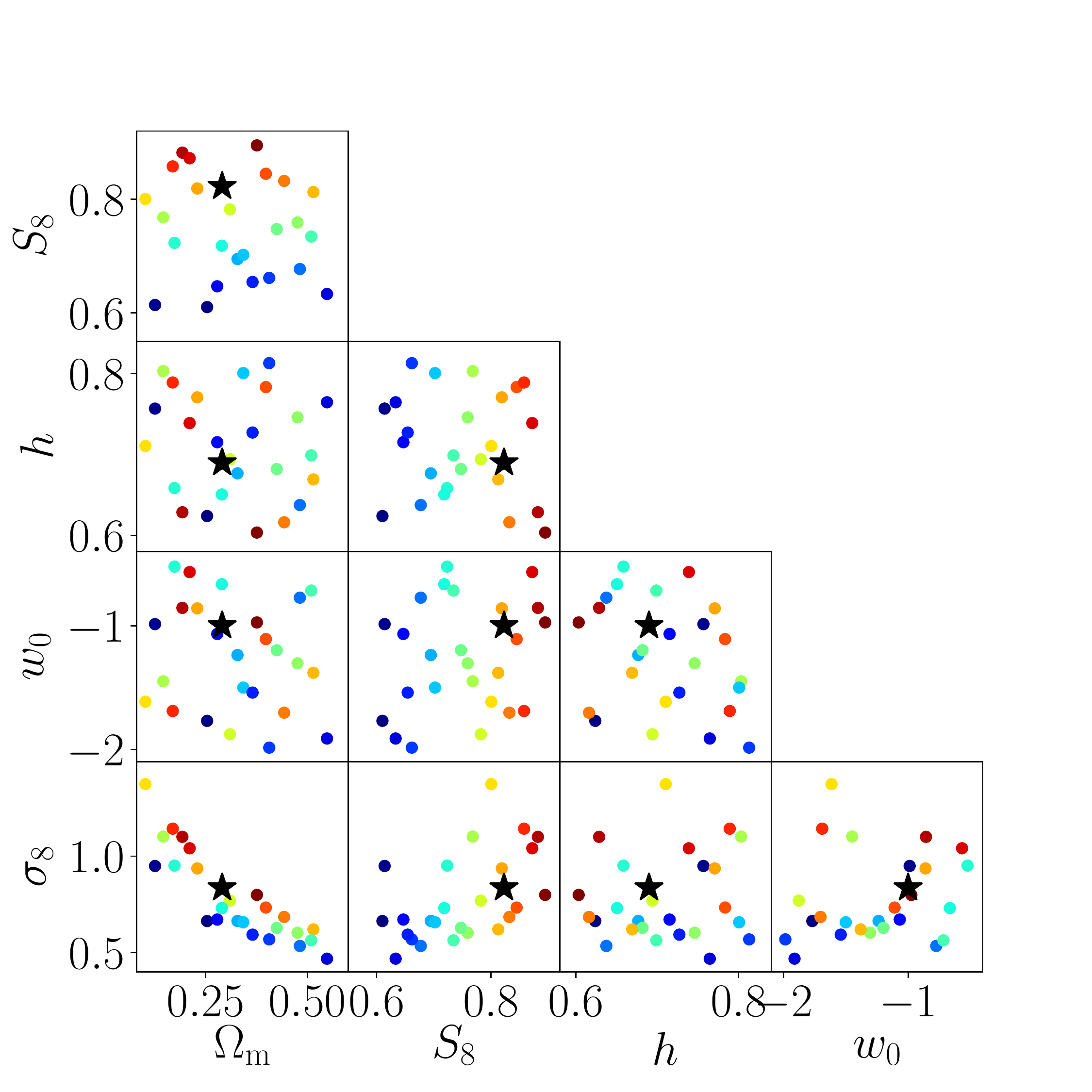} \\
\caption{
The 26 distinct cosmologies in cosmo-SLICS, colour-coded by $S_8$ except for the fiducial cosmology which is denoted by the black star. The upper three rows show the relatively even spread of nodes with respect to $\Omega_{\rm m}$, $S_8$, $h$ and $w_0$ owing to the Latin hypercube sampling of these parameters. The lowest row shows these cosmologies translated into the $\sigma_8$ space. All nodes share the same value of the baryon energy density $\Omega_{\rm b}=0.0473$, spectral index $n_{\rm s} =0.969$, and neutrino energy density $\Omega_\nu = 0$. The fiducial cosmology is the only $\Lambda$CDM model $(w_0=-1)$ and is used for all realisations in SLICS.} \label{fig:cosmo-SLICS}
\end{figure}

The SLICS (Scinet LIghtcone Cosmological Simulations) and the subsequent cosmo-SLICS were created using the gravity solver CUBEP$^3$M \citep{harnois-deraps/etal:2013}. These dark-matter-only simulations have a lightcone area, $A_{\rm lc}=100 \, \rm{deg}^2$, box size, $L_{\rm box} = 505 \, {\rm{Mpc}}/h$ per side, with each containing $1536^3$ particles. SLICS consists of more than 600 independent realisations all with the same fiducial set of cosmological parameters and contains particles of mass $2.88 \times 10^9 M_{\odot}/h$. 
In contrast, the cosmo-SLICS suite consists of 1,250 $N$-body simulations divided between 26 distinct cosmologies, varying in 
$\Omega_{\rm m}$, $S_8 = \sigma_8 \sqrt{ \Omega_{\rm m}/0.3 }$, the Hubble parameter $h$ (in units of 100 $\rm{Mpc}/\rm{kms}^{-1}$) and the dark energy equation of state parameter $w_0$. The particle masses in cosmo-SLICS vary with $\Omega_{\rm m}$ and $h$ in the range $[1.42,7.63]\times 10^9 M_{\odot}$.
The 26 distinct input cosmologies, or ``nodes", occupy the 4-dimensional parameter space in a Latin hypercube configuration, such that all values of the parameters are unique to each node. This is to facilitate a fairly even sampling of the parameter volume. The distribution of the nodes is presented in Figure \ref{fig:cosmo-SLICS}.

For each cosmology in cosmo-SLICS, there are 50 realisations: 25 pairs with initial conditions carefully chosen such that the initial power spectra of each pair average to a close approximation of the theoretical prediction \citep{harnois-deraps/etal:2019}. This pairing strategy, inspired by the methodology of \cite{angulo/pontzen:2016}, suppresses the impact of cosmic variance on the expectation value of a given statistic. In turn this makes the accurate estimation of cosmological statistics feasible from a tractably-small number of numerical simulations, and for a broad range of cosmological parameters. 

In this work we are investigating the constraining power of the lensing PDF for realistic current and future survey specifications. To that end, we utilise SLICS and cosmo-SLICS which are tailored to match the number density, galaxy shape noise, and redshift distributions of the most recent data release from the Kilo-Degree Survey \citep[][hereafter referred to as ``KiDS-1000"]{giblin/etal:2021, hildebrandt/etal:2021}. These catalogues contain shear and photometric redshift estimates for 21 million galaxies spanning 1006 square degrees of sky coverage with an effective galaxy number density of 6.2 galaxies per square arcminute. In this work, we mimic the recent cosmic shear analysis of this data \citep{asgari/etal:2021} by using the same five tomographic redshift bins bordered by $[0.1, 0.3, 0.5, 0.7, 0.9, 1.2]$. We also employ a version of the SLICS and cosmo-SLICS with five times higher number density, 30.5 galaxies per square arcminute, in order to simulate upcoming lensing data from the Vera Rubin Legacy Survey of Space and Time \citep{lsst/etal:2012}. In our ``LSST" analysis, we use the same tomographic redshift binning and galaxy shape noise as with the KiDS-1000 version of the simulations. 

\new{Our simulations differ notably from the {\sc MassiveNuS} \citep{liu/etal:2018} employed by \citet{liu/madhavacheril:2019} to model lensing PDFs. Namely, they sample a 3D $\Lambda$CDM parameter space defined by $\Omega_{\rm m}$, the sum of the neutrino masses $\sum M_\nu$, and the primordial power spectrum amplitude $A_{\rm s}$, with many more nodes (101) and realisations (10,000 per node), but at the cost of a smaller lightcone area (12.25 deg$^2$) and fewer particles ($1024^3$). Whilst we both simulate an LSST-like survey, \citet{liu/madhavacheril:2019} assume a slightly larger galaxy number density ($\sim 50 \, \rm{arcmin}^{-2}$) and higher tomographic redshift binning (five linear-spaced bins out to $z=2.5$). This means that our estimates of the relative power of lensing PDFs to two-point statistics are produced under different conditions and are not directly comparable. }


\section{Methodology} \label{sec:method}

In order to build intuition on the cosmological constraining power of the lensing PDF and to verify our pipeline, we begin in Sections \ref{subsec:method_gaussian}--\ref{subsec:method_lognormal} with performing mock cosmic shear analyses using this relatively new statistic measured from Gaussian and lognormal fields generated with the {\sc healpy} \citep{zonca:2019} and {\sc FLASK} \citep{xavier/etal:2016} software packages respectively. We compare the lensing PDF constraints with those from the standard two-point lensing power spectrum given by, 

\begin{equation} \label{eqn:Cell}
C^{ij}(\ell) = \int_0^{\chi_{\rm H}} \text{d} \chi \, \frac{q_i(\chi)q_j(\chi)}{f_{\rm{K}}(\chi)^2} \, P_\delta \left( k=\frac{[\ell+1/2]}{f_{\rm{K}}(\chi)},\chi \right) \,,
\end{equation}

\noindent where $\chi$ is the comoving radial distance to the horizon, $f_{\rm{K}}(\chi)$ is the comoving angular diameter distance (equal to $\chi$ in a spatially flat universe), and $P_\delta(k,\chi)$ is the matter power spectrum as a function of the wavenumber, $k$, the Fourier conjugate of $\chi$. The relationship between the wavenumber and the multipole $\ell$ shown here \citep[particularly noting the $+1/2$ term which differs from the corresponding equation in, for example,][]{bartelmann/schneider:2001} is an expression of the flat-sky first-order extended Limber approximation \citep{loverde/afshordi:2008,kilbinger/etal:2017}. The integral is taken from the observer at $\chi=0$ to the horizon, $\chi_{\rm{H}}$.

The lensing efficiency, $q_i(\chi)$, where $i$ denotes the redshift bin index, is given by

\begin{equation} \label{eqn:lensing_efficiency}
q_i(\chi) = \frac{3 H_0^2 \Omega_{\rm m}}{2c^2} \frac{f_{\rm{K}}(\chi)}{a(\chi)}\int_\chi^{\chi_{\rm H}}\, \text{d} \chi^\prime\ n_i(\chi^\prime) 
\frac{f_{\rm{K}}(\chi^\prime-\chi)}{f_{\rm{K}}(\chi^\prime)}.
\end{equation}
\noindent Here $a$ is the scale factor, $n_i(\chi)$ is the galaxy number density as a function of comoving radial distance within redshift bin $i$, $H_0$ is the present-day Hubble parameter and $c$ is the speed of light.

In the case of Gaussian maps, the lensing power spectrum fully captures the cosmological information in the field and we expect the size of the lensing PDF constraints to be larger than, or at best equal in size to, those from the conventional statistic. In the lognormal scenario however, where the non-Gaussianity renders some information beyond the reach of the two-point power spectrum, the relative performance of the lensing PDF is less certain, and could in principle outperform the former statistic. 

In addition to facilitating these sanity-check tests, {\sc healpy} and {\sc FLASK} have the added benefit of providing lensing predictions for arbitrary choices of cosmological parameters (within a certain range). Thus, these packages provide the freedom to perform a high-resolution grid-based likelihood analysis. In our tests with the full $N$-body simulated lensing maps, discussed in Section \ref{subsec:method_nbody}, the more accurate reconstruction of the matter clustering in our Universe means that the computational expense of producing these for each cosmology prohibits a grid-based likelihood evaluation. We therefore switch to an MCMC sampler twinned with a Gaussian process emulator trained on the simulated measurements. 
\\
\\
\subsection{Gaussian fields} \label{subsec:method_gaussian}
In the case of a Gaussian density field, we expect the lensing PDF to perform equally well or worse than the optimal statistic, the two-point lensing power spectrum, in constraining cosmological parameters. This test therefore serves as a reliable way to validate our pipeline and gain intuition as to the information contained in this novel lensing probe. 

Using {\sc healpy} \citep{zonca:2019}, we generate full-sky noise-free Gaussian-density HEALPix \citep{gorski/etal:2005} convergence maps with NSIDE 256. These are produced from input lensing power spectra, $C(\ell)$, generated using the {\sc Nicaea} code from \citet{kilbinger/etal:2009} with the HaloFit model from \cite{smith/etal:2003}. For the galaxy redshift distribution we arbitrarily set all sources to be at $z=1$ and measure our statistics for only one broad redshift bin. We produce input $C(\ell)$s for 40,000 different cosmologies, varying $\Omega_{\rm m}$ and $\sigma_8$ on a grid of 200$\times$200 dimensionality. The range of these parameters is selected to cover concurrent cosmic shear constraints \citep{asgari/etal:2021}. {We also adjust the dark energy density parameter to maintain a flat $\Lambda$CDM cosmology, $\Omega_{\Lambda}=1-\Omega_{\rm m}$.} All other cosmological parameters remain fixed. We produce 10 realisations of the maps at each cosmology by varying the random seed, and we average the PDFs measured from these to get smooth statistics for each cosmology. The same random seeds are used for all of the grid cosmologies to eliminate the effect of cosmic variance between the different models. 

In this work, we are interested in measuring the improvements in cosmological precision when the lensing PDF is measured over multiple angular scales, as observed by \cite{boyle/etal:2021} \new{and \citep{kratochvil/etal:2012}}. For that reason, we measure the PDF of the convergence maps with different levels of smoothing applied to the field. Hence, we produce one set of maps without any smoothing and two more sets with increasing degrees of smoothing applied.

 In the no-smoothing case, the only limitation on the scale-dependent information comes from the resolution of the HEALPix maps, 13.7 arcmin. We note that concurrent cosmic shear surveys use scales as low as $\sim 1$ arcminute. However, we find that maps with our chosen resolution, as well as being computationally cheap to produce, are sufficient to demonstrate the difference in cosmological information contained in the lensing PDF and two-point statistics. The other two sets of maps are smoothed using mean-zero Gaussian kernels with two different values of the standard deviation, $\sigma_{\rm s} = 30$ arcmin and 60 arcmin. These values are selected simply on account of being larger than the map resolution, thereby yielding a detectable amount of smoothing, but smaller than the upper range of scales typically used in cosmic shear studies ($\sim 300$ arcmin). Apart from these criteria the values used are somewhat arbitrary and are unlikely to be optimal choices, but in practice they yield noticeable breaking of the degeneracy in the lensing PDF constraints in the $\Omega_{\rm m}-\sigma_8$ plane.

In order to evaluate the covariance of the lensing PDF - a necessary ingredient of any likelihood analysis - we pick the approximate centre of our $\Omega_{\rm m}-\sigma_8$ grid to serve as the ``data" cosmology and generate $N=1000$ map realisations by varying the random seed. The covariance is calculated from the subsequent PDFs via,
\begin{equation} \label{eqn:CovMat}
\Sigma_{\rm s}(\kappa_i, \kappa_j) = \sum_k^{N}  \frac{ \left( {p_{\rm s}^k(\kappa_i)} - \overline{p_{\rm s}}(\kappa_i) \right) \left( {p_{\rm s}^k(\kappa_j)} -  \overline{p_{\rm s}}(\kappa_j) \right)}{N-1} \,,
\end{equation}
where $p_{\rm s}^k(\kappa_i)$ denotes the $i$'th bin in convergence $\kappa$, of PDF realisation $k$ measured for a given level of smoothing ${\rm s}$, and $\overline{p_{\rm s}}(\kappa_i)$ denotes its ensemble average over the $N$ realisations. We also measure the combined covariance of the PDFs across the different smoothing levels by concatenating the PDFs for each realisation $k$. 

\begin{figure}
\centering
\includegraphics[width=0.49\textwidth]{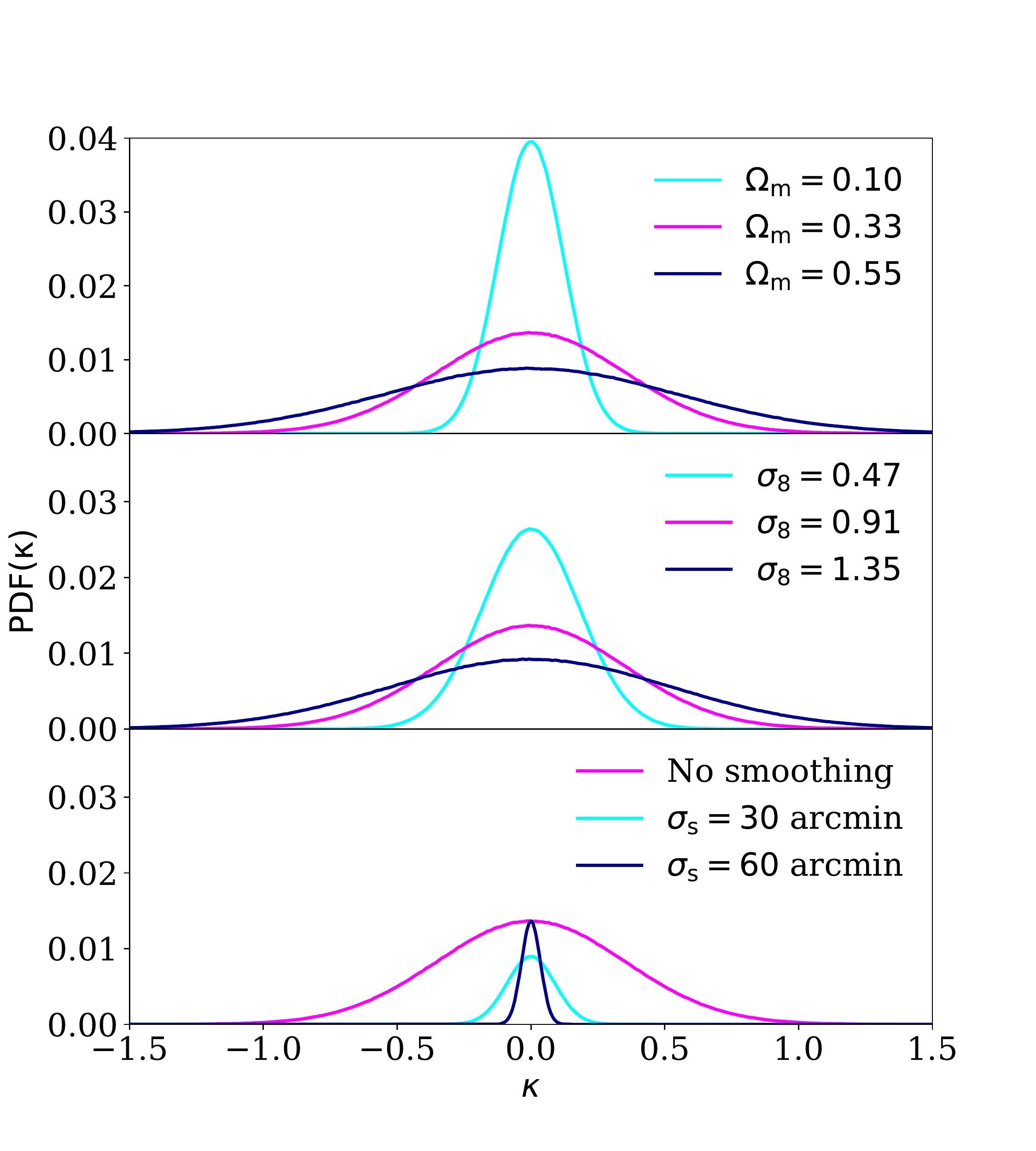} 
\caption{Convergence PDFs measured from full-sky Gaussian-density HEALPix maps generated using {\sc healpy}, varying in $\Omega_{\rm m}$ (upper panel), $\sigma_8$ (middle panel) and the amount of smoothing (lower panel). The upper two panels correspond to maps with no smoothing, and the magenta line on each panel is the same, corresponding to the no-smoothing case with intermediate values of the cosmological parameters.} \label{fig:PDFs_HP}
\end{figure}

The variation in the lensing PDFs with $\Omega_{\rm m}$, $\sigma_8$, and the level of applied smoothing measured from the Gaussian {\sc healpy} maps is presented in Figure \ref{fig:PDFs_HP}. As $\Omega_{\rm m}$ and $\sigma_8$ increase independently in the absence of smoothing (upper two panels), intuitively, we see the lensing PDF broadening. Increases in these parameters correspond to stronger matter clustering and hence greater abundances of over- and underdense regions. This manifests in a greater number of pixels with more positive or more negative values in our convergence maps. 

As we increase the level of smoothing with the cosmological parameters held fixed (lower panel of Fig.~\ref{fig:PDFs_HP}), the lensing PDF narrows. This stems from the suppression of small-scale perturbation modes with the wider Gaussian kernel, which leads to a distribution of pixel values with decreased variance. In practice, this alters the physical scale from which cosmological information is being garnered, with less smoothing giving greater weight to information from smaller scales.

The decreasing variance of the lensing PDFs with increasing smoothing mean that one must adopt a smoothing-scale-specific convergence binning scheme for the likelihood analysis. Excessive sampling of the PDF tails, where ${\rm PDF}(\kappa) \simeq 0$, leads to a covariance which cannot be accurately inverted, whilst using a large number of bins in the data vector produces a noisy covariance. For these reasons we adopt a binning scheme which: 1) samples the PDF with four evenly-spaced $\kappa$ bins symmetrically about the origin, and 2) contains approximately 2$\sigma$ of the PDF volume, where the standard deviation is inferred by eye from the data cosmology PDFs separately for each smoothing scale. In practice with our {\sc healpy} Gaussian maps, this leads to sampling the convergence values, for all cosmologies, in the range $\pm 0.60$, $\pm 0.1$ and $\pm 0.06$ for the no-smoothing, $\sigma_{\rm s} = 30$ arcmin and $\sigma_{\rm s} = 60$ arcmin cases respectively. 

Recognising that this binning scheme is unlikely to be an optimal choice, we experimented with widening the $\kappa$ range used for each smoothing scale, and with adding extra broad $\kappa$ bins containing the extreme pixel values from the PDF tails. We found that these tests resulted in noisier likelihood surfaces without noticeable improvements to parameter constraints. This suggests that any increases in information garnered from these extreme pixels is minimal and/or subdominant to the numerical instabilities introduced by these pixels to the inversion of the covariance. The diminishing returns with wider PDF binning could in principle be overcome using transformations of these statistics or through compression (for example, using principal components). We find, however, that our approach to the binning results in smooth and consistent PDF cosmological constraints which adhere to expectations in the case of the Gaussian maps. 

Our intention in this work is to compare the constraining power of the lensing PDF with standard two-point lensing statistics. Therefore, we also measure the output $C(\ell)$s from the un-smoothed convergence maps using the {\sc healpy} spherical harmonic transform function \textit{anafast}, and average them over the ten realisations per cosmology in the same way as the lensing PDFs. We use these output $C(\ell)$s as the measurements entering into our likelihood evaluation, rather than the original input $C(\ell)$s from {\sc Nicaea}, because it is important to have the same scale cuts imposed by the resolution of the HEALPix maps on the power spectra as on the PDFs, in order to facilitate a fair comparison. We measure the output $C(\ell)$s in fifteen logarithmically-spaced bins with $\ell \in [1, 3 \times {\rm NSIDE}]$, with the upper limit imposed by the map resolution (${\rm NSIDE = 256}$). The lensing power spectrum covariance is computed from 1000 realisations in the same way as the PDFs, following equation \ref{eqn:CovMat} and simply substituting the PDF for $C(\ell)$. 

\begin{figure*}
\centering
\includegraphics[width=0.49\textwidth]{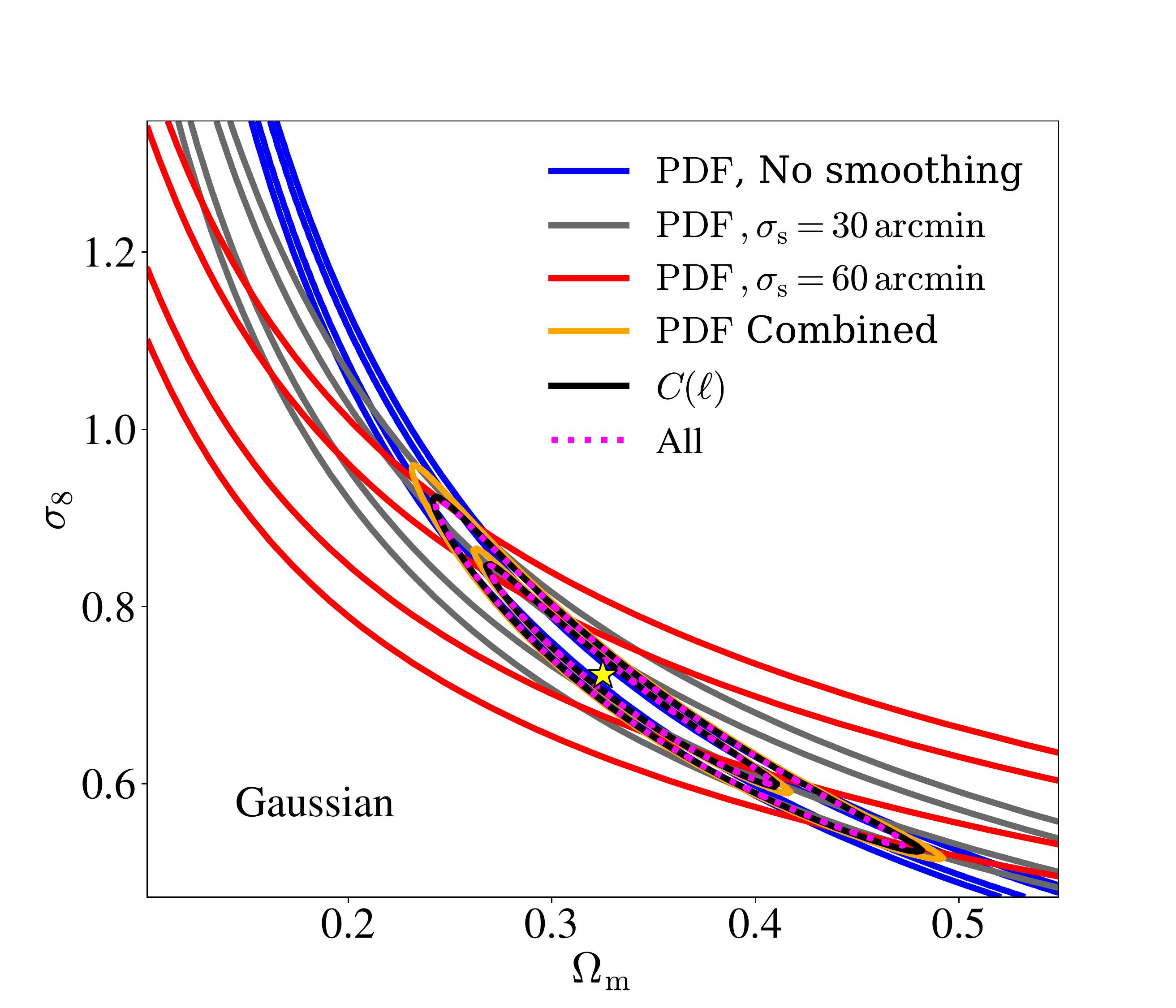} 
\includegraphics[width=0.49\textwidth]{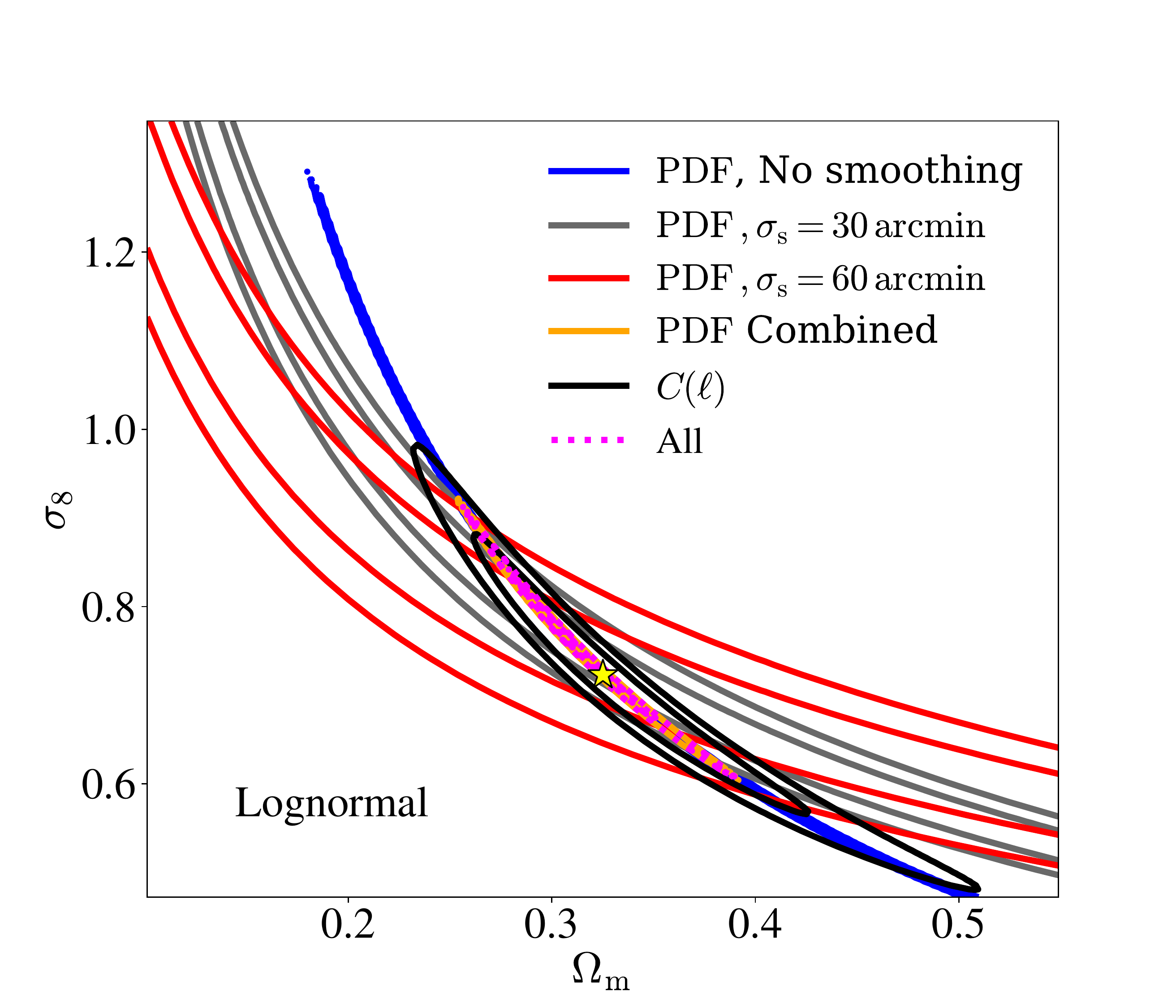} 
\caption{Cosmological constraints for a Gaussian (left) and lognormal (right) mock density field with an input cosmology designated by the star. The PDFs of convergence maps with increasing levels of smoothing (no smoothing, and smoothing scales of 30 arcmin and 60 arcmin) give rise to the blue, grey, and red contours respectively. The combination of the PDFs gives the orange contour. The constraints from the standard $C(\ell)$ lensing power spectrum are shown in black, and the combination of all statistics gives the dashed magenta contour. Note that the `All' contour is superimposed with the $C(\ell)$ (black) contour in the left panel, and with the PDF combined (orange) contour in the right panel.} \label{fig:constraints_gauss_field}
\end{figure*}

We evaluate the grid-based Bayesian posterior probability of cosmological parameter configurations, $\boldsymbol{\pi}$, given the data $\boldsymbol{d}$, itself averaged from the 1000 realisations used to compute the covariance, as
\begin{equation}
p(\boldsymbol{\pi}|\boldsymbol{d}) = \frac{\mathcal{L}(\boldsymbol{d}|\boldsymbol{\pi})p(\boldsymbol{\pi})}{E} \,,
\end{equation}
where $\mathcal{L}(\boldsymbol{d}|\boldsymbol{\pi})$ denotes the likelihood, $p(\boldsymbol{\pi})$ is the prior probability of cosmology $\boldsymbol{\pi}$ and $E$ is the evidence, which normalises the integral of the posterior over all possible values of $\boldsymbol{\pi}$ to unity. Here we adopt uniform priors on $\Omega_{\rm m}$ and $\sigma_8$ which cover the extent of the 2D grid pictured in Figure \ref{fig:constraints_gauss_field}. The likelihood, or the probability of observing the data given a certain cosmology is the truth, is defined as
\begin{equation} \label{eqn:Lhd}
\mathcal{L}(\boldsymbol{d}|\boldsymbol{\pi}) \propto  \exp \left( -\frac{1}{2} \left[\boldsymbol{d} - \boldsymbol{m}(\boldsymbol{\pi}) \right]^\intercal \Sigma^{-1} \left[\boldsymbol{d} - \boldsymbol{m}(\boldsymbol{\pi}) \right] \right) \,,
\end{equation}  
where $\boldsymbol{m}(\boldsymbol{\pi})$, in our case, represents our model for the lensing PDF or power spectrum as a function of cosmology. $\Sigma^{-1}$ is the inverse covariance matrix (Eq.~\ref{eqn:CovMat}) modulo two scaling factors. The first, is the (in this case small) \citet{hartlap/etal:2007} correction factor which accounts for biases incurred from inverting covariance matrices which have been estimated from a finite number of samples. The second rescales the covariance so as to simulate a 500 square-degree survey, rather than the full-sky area over which the covariance is calculated. This simply facilitates cosmological constraints which are comparable in size to concurrent weak lensing surveys and extend over an appreciable fraction of the 2D cosmology grid. 

The left-hand panel of Figure \ref{fig:constraints_gauss_field} presents the posterior probability constraints from the $C(\ell)$ (in black), the individual lensing PDFs at each smoothing level (blue, grey, red), and the combined PDF constraints from the three smoothing levels (orange), for the Gaussian field with true cosmology denoted by the star. Also presented, superimposed with the $C(\ell)$ contours, are the combined constraints from all statistics - $C(\ell)$ and the PDFs - in dashed magenta. Immediately apparent is the anticlockwise rotation in the lensing PDF constraints as progressively more smoothing is applied (blue, to grey, to red contours respectively). This manifests in a breaking of the $\Omega_{\rm m}-\sigma_8$ degeneracy when the PDF constraints are combined, leading to a considerable improvement in precision for both parameters, consistent with observations by \citet{boyle/etal:2021}. 

Intuitively we see that of the three individual PDFs, the no-smoothing case leads to the tightest constraints in terms of area. This is because this measurement probes an angular scale akin to the resolution of the maps, $\sim 13.7$ arcmin, {and contains more perturbation modes and smaller sample variance} than  the smoothed PDFs. However, the rotation of the contours as smoothing is introduced and increased implies that larger-scale features help to better constrain the amplitude of density fluctuations, $\sigma_8$. Whilst tests with the different PDF combinations reveal that most of the constraining power comes from the un-smoothed case, all three of the smoothing levels considered here contribute to reducing the combined contour area. This is because the variance of the density field is scale-dependent, as set by the shape of the power spectrum. It is necessary to have PDFs at several different scales to capture the cosmological information encoded in this shape. 

Despite the improvements in precision offered by the combinations of PDFs with different smoothing levels, in this particular test with a Gaussian field, we expect and indeed find the standard two-point measurement (black contour) to reign supreme; we see that the combinations of multi-scale PDFs approach but do not exceed the precision of the $C(\ell)$. Furthermore, the dashed magenta contours, produced from the combination of all statistics, are superimposed with those of the $C(\ell)$, further verifying that the PDFs do not contain any information not already captured by the optimal standard two-point probe.  These findings are an encouraging validation of our pipeline, and a promising suggestion of the gains in constraining power the lensing PDF could facilitate when applied to non-Gaussian fields.

\subsection{Lognormal fields} \label{subsec:method_lognormal}

Having gained intuition on the impact of combining multi-scale PDFs and having verified our methodology with the PDF-$C(\ell)$ comparison for a Gaussian field, we proceed to investigate the relative strengths of these statistics with fields harbouring non-Gaussianities. In this case we use lognormal fields, these being a reasonably good approximation of the late-time density field in the real Universe. With these fields it is possible in principle for the PDF measured on different scales to extract more cosmological information than the standard two-point probe, given the latter is optimal only for Gaussian fields. {Hence this test serves as a means of qualitatively understanding the difference in information contained in the lensing PDFs and two-point statistics.

We perform a near-identical analysis as in Section \ref{subsec:method_gaussian}, but this time we generate lognormal convergence {\sc HEALPix} maps using {\sc FLASK} software \citep{xavier/etal:2016}. We use the same smoothing levels (none, $\sigma_{\rm s}=30$ and 60 arcmin) as before, but we adjust the $\kappa$ binning to account for the fact that the PDFs for lognormal fields are no longer symmetric about the origin. Nevertheless we continue to use the same approach to binning the convergence maps, aiming for as wide as possible coverage of the range of pixel values with four $\kappa$ bins, without excessively sampling the tails measured from the data cosmology which leads to noisier likelihood evaluations. Four bins is sufficient to probe up to the fourth-order moments of the density field, a significant improvement beyond the lensing power spectrum which samples only up to second-order. \new{Probing higher order moments with additional bins is unlikely to improve constraints significantly given that the second and third order moments of the field have been shown to be the dominant sources of cosmological information \citep{boyle/etal:2021}.} We continue to use the same multipole binning for the $C(\ell)$.

The cosmological constraints for the lognormal field from the lensing PDFs, their combination, the $C(\ell)$, and finally the combination of all statistics, are displayed in the right-hand panel of Figure \ref{fig:constraints_gauss_field}. We observe a similar rotation of the lensing PDF contours when smoothing is increased as with the Gaussian field shown in the left panel of this figure, and a corresponding considerable improvement  in the precision when the PDFs are combined. A marked difference to the Gaussian-field analysis is the fact that the combined-PDF now outperforms the $C(\ell)$, producing marginalised $\Omega_{\rm m}$ and $\sigma_8$ constraints which are $\sim 50\%$ and $\sim 35\%$ tighter, respectively. This finding confirms our expectation that the lensing PDF is privy to cosmological information which is beyond the reach of the standard two-point functions due to the presence of non-Gaussianities in the field. This is also only achievable by combining PDFs of several different smoothing scales, for the reasons discussed in the previous subsection.

In further contrast to the results obtained with the Gaussian field, the dashed magenta contour - corresponding to the combination of all statistics - is now superimposed with the orange (PDF combined) contour, rather than the black constraints from the $C(\ell)$. This means that the PDFs measured from the lognormal field over various scales capture all of the information within the $C(\ell)$, such that including the two-point statistic in the concatenated PDF vector in the likelihood contributes no extra constraining power. In the case of the Gaussian field this situation was reversed - the $C(\ell)$ harnessed all information available in the field. This change of course arises from the introduction of non-Gaussianities to the field; whereas the $C(\ell)$ is limited to sampling information up to second order, the lensing PDFs successfully extract information up to this limit and beyond.

Comparing the two black contours between the left and the right-hand panels of \ref{fig:constraints_gauss_field}, it is also noticeable that the $C({\ell})$ constraints are worse when the field is lognormal (right). This agrees with the expectation that the $C({\ell})$s are insufficient to capture all of the cosmological information when the field is non-Gaussian.

\subsection{$N$-body simulated fields} \label{subsec:method_nbody}

\subsubsection{KiDS-1000 analysis} \label{subsubsec:meth_k1000}

We now proceed to our cosmic shear analysis using full $N$-body numerical simulations - the SLICS and cosmo-SLICS - which are designed to faithfully reconstruct many of the specifications of concurrent and future weak lensing surveys. We begin with the simulation products tailored to match the shape noise, number density, and galaxy redshift distributions of the most-recent KiDS data release, KiDS-1000 \citep{giblin/etal:2021}. Reproducing these qualities of the data in the SLICS and cosmo-SLICS has two effects. First of all, the $N$-body simulations differ significantly from the Gaussian and lognormal fields considered previously, and different analysis choices, discussed further in this section, need to be made to reflect this. Secondly, the cosmic shear experiment presented here is notably more realistic, albeit one which is still free of systematics.

To reproduce the shape noise properties of the data in the simulations, we generate Gaussian-distributed random galaxy ellipticities truncated to $\in [-1,1]$ with standard deviations between 0.254 and 0.273 depending on the redshift bin\footnote{We refer the interested reader specifically to Table 1 of \citet{giblin/etal:2021}.}. We then calculate observed galaxy ellipticities as \citep{seitz/schneider:1997},
\begin{equation} \label{eqn:e_obs}
\boldsymbol{e}_{\rm obs} = \frac{\boldsymbol{g} + \boldsymbol{e}_{\rm int}}{1+ \boldsymbol{g}^*e_{\rm int}} \,,
\end{equation}
where $\boldsymbol{g} = g_1 + ig_2$ is the complex reduced shear which in weak lensing, where the convergence $\kappa$ is small, is approximately equal to the shear $\boldsymbol{\gamma}$ itself: $\boldsymbol{g} = \boldsymbol{\gamma}/(1-\kappa) \simeq \boldsymbol{\gamma}$. The randomly-generated intrinsic ellipticity is denoted by $\boldsymbol{e}_{\rm int}$ and the complex conjugate by $*$. 

Whilst versions of the SLICS and cosmo-SLICS which replicate the geometry of the KiDS-1000 footprint do exist \citep{harnois-deraps/etal:2021}, as this is a proof-of-concept study, we do contend with the systematics that can potentially arise from survey masking \citep[see for example][]{giblin/etal:2018}. Therefore we use mocks in which the galaxies are randomly distributed on the angular plane and leave the impact of masking for future study. 

The SLICS and cosmo-SLICS, at $10\times 10 \, \rm{deg}^2$ per map, are significantly smaller than the full-sky Gaussian and lognormal HEALPix maps considered in the previous sections. This means it is computationally feasible to increase the angular resolution in this analysis and to use a flat-sky mass reconstruction approach instead of the spherical harmonic methodology employed by {\sc healpy} and {\sc FLASK}. Hence, from catalogues of the observed galaxy ellipticities and angular positions, we produce noisy convergence maps following the conventional \citet{kaiser/squires:1993} approach. This involves converting our galaxy ellipticity catalogues into two-dimensional ellipticity maps with dimensionality 600$\times$600 pixels (1 arcmin$^2$ per pixel), using the angular positions of galaxies to project them onto the grid. Where more than one galaxy is projected onto a given pixel their ellipticities are averaged. The ellipticity maps are smoothed with a Gaussian kernel before the \citet{kaiser/squires:1993} methodology is employed to produce convergence maps.

As in the previous sections, we produce these convergence maps with different levels of smoothing. However, the increased angular resolution compared to the Gaussian/lognormal test cases means that we are now free to sample smaller-scale cosmological information through the use of narrower Gaussian smoothing kernels. Hence, we reduce the standard deviations from the previous values of 30 and 60 arcmin to $\sigma_{\rm s} = $2.2, 6.6 and 13.2 arcmin. The intermediate smoothing scale was selected on account of this being identified in previous analyses as an appropriate scale for suppressing noise without sacrificing too much of the cosmological information contained on small scales \citep{giblin/etal:2018, vanwaerbeke:2000}. For the lower and higher smoothing scales we arbitrarily selected values three times smaller and two times larger than the intermediate scale respectively. 

We do not produce maps without any smoothing, as we did before, since the introduction of shape noise to the mocks creates localised spikes in the ellipticity distributions which manifest in systematic contributions to the convergence maps over a range of scales. It is prudent to apply some smoothing to suppress this effect. We also measured PDFs using the largest of the smoothing scales employed with the Gaussian and lognormal maps, $\sigma_{\rm s} = $60 arcmin, but found this added no noticeable improvement to the combined constraining power of the other three smaller scales. We therefore discarded the results produced with this level of smoothing. 

  \begin{figure*}
\centering
 \includegraphics[width=\textwidth]{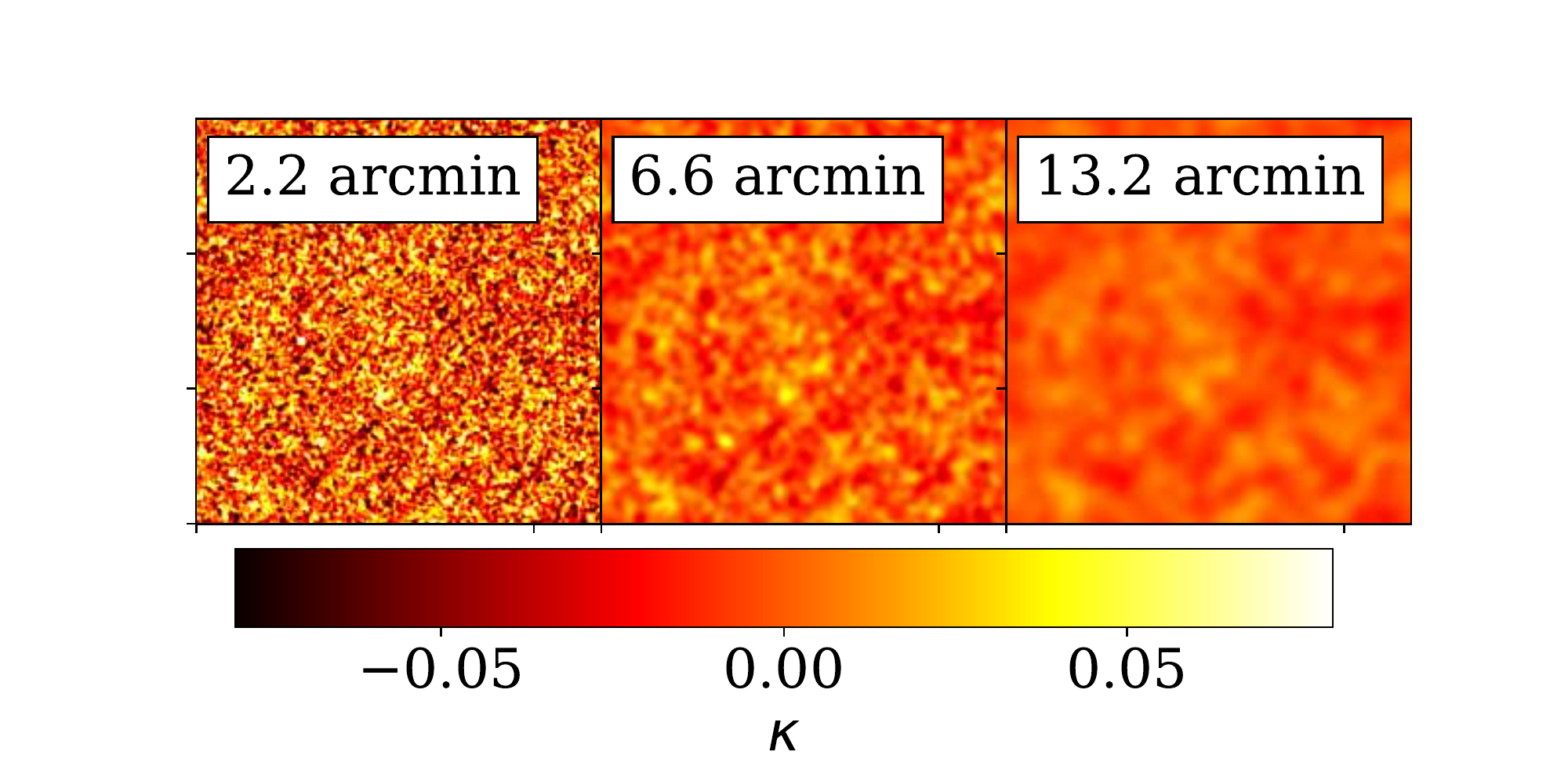} \\
\caption{Convergence maps from the same fiducial cosmology lightcone of the KiDS-1000 version of cosmo-SLICS with the three levels of smoothing used in our mock cosmic shear analyses: 2.2, 6.6 and 13.2 arcminutes (left to right panels respectively). Each map measures $10 \times 10 \, \rm{deg}^2$ and corresponds to the highest tomographic redshift bin $(0.9<z<1.2)$.} \label{fig:kappa_maps} 
\end{figure*}

Figure \ref{fig:kappa_maps} illustrates the impact of different smoothing scales on the convergence maps for a given lightcone and realisation of the galaxy shape noise selected at random from the cosmo-SLICS fiducial cosmology maps in the highest tomographic redshift bin $(0.9<z<1.2)$. As the smoothing level is decreased, progressively smaller projected matter structures manifest in the maps, potentially providing extra cosmological information but at the expense of increasingly prevalent noise features. This effect establishes a careful balance which must be struck in order to effectively use the lensing PDFs to constrain cosmological parameters: one should aim to include as much small-scale information as possible whilst keeping the noise levels sufficiently low to achieve smooth and unbiased cosmological constraints.

In addition to measuring the lensing PDFs in each of the five tomographic redshift bins used in the KiDS-1000 cosmic shear analysis \citep{asgari/etal:2021}, henceforth referred to as the ``auto-PDFs", we also measure the ``cross-PDFs" from the ten possible redshift-bin pair combinations. We undertake this in the same manner as \citet{harnois-deraps/etal:2021} and \citet{martinet/etal:2021}: by concatenating two ellipticity catalogues from the same lightcone but different redshift bins before performing the \citet{kaiser/squires:1993} mass reconstruction to produce a ``cross-redshift" convergence map, from which cross-PDFs are measured. In the case of adjacent redshift ranges this is equivalent to merging the two bins, but for non-adjacent bins the subsequent map contains a combination of projected structures along the line of sight, with the contributions from each redshift range varying depending on the bins which are concatenated. The extra information which the cross-PDFs add to the posterior probability evaluation is akin to the information contained in the cross-correlations defined by equation \ref{eqn:Cell}, where the redshift bin indices differ, $i \neq j$. 

Including tomography here represents a further deviation from the Gaussian and lognormal test cases in Sects.~\ref{subsec:method_gaussian}-\ref{subsec:method_lognormal}, in which only a single broad redshift bin was used. Like the inclusion of galaxy shape noise, this change was made to make this analysis as close as possible to the recent KiDS-1000 cosmic shear study \citep{asgari/etal:2021} save for the introduction of observational systematics. We note that in addition to the redshift bin pair combinations, one could also include PDFs calculated from triplet, quadruplet, and quintet combinations of redshift bins, with the latter being equivalent to having only one broad redshift bin. These extra measurements may add further constraining power to the lensing PDF approach. Owing to the increased computational time required to produce these measurements from the simulations, however, we leave this avenue of investigation for future work.

When measuring cosmological statistics from noisy convergence maps, for example when working with the real data, it is sensible to boost the signal by dividing the observed projected density by one quantifying the noise present in the field. Hence, we also produce noise-only convergence maps from catalogues containing galaxies with only random intrinsic ellipticities and zero shear. This follows the same procedure as with the catalogues which also contain the signal, using the same three smoothing levels. In the case of the cross-redshift maps, this means concatenating catalogues containing only random ellipticities. We calculate the variance of the pixel values from 715 noise-only map realisations for each redshift bin combination and smoothing scale, in order to quantify the noise level in the field. We then divide our ``signal$+$noise" convergence maps by the noise estimate specific to each redshift range and smoothing scale to produce signal-to-noise ratio (SNR) maps, from which we measure the lensing PDFs.

It is particularly important to include shape noise when simulating a realistic cosmic shear analysis involving the lensing PDF because the width of this statistic is sensitive to the distribution of observed ellipticities. This can be seen from the relationship between the galaxy ellipticity dispersion, $\sigma_{\rm e}$, and the noise dispersion in the convergence maps given by \citep{kaiser/squires:1993, waerbeke:2000},
\begin{equation} \label{eqn:kappa_noise}
\sigma_\kappa^2 = \frac{\sigma_{\rm e}^2}{2} \frac{1}{2\pi\sigma_{\rm s}^2 n_{\rm gal}}.
\end{equation} 
As the intrinsic galaxy shape noise is increased, for a fixed smoothing scale, $\sigma_{\rm s}$, and galaxy number density, $n_{\rm gal}$, the dispersion of convergence pixel values similarly rises, which results in a wider lensing PDF in convergence space. In order to suppress any biases arising in the shape of the PDFs from particular realisations of the random shape noise, we generate 20 different galaxy shape noise realisations for each of the 50 lightcones per cosmology and redshift bin. The lensing PDFs measured from the maps produced for these realisations are subsequently averaged. This allows for the shape noise properties of the KiDS-1000 data to be reproduced in the simulations whilst mitigating the impacts of shot noise in constructing our model.

  \begin{figure*}
\centering
\includegraphics[width=\textwidth]{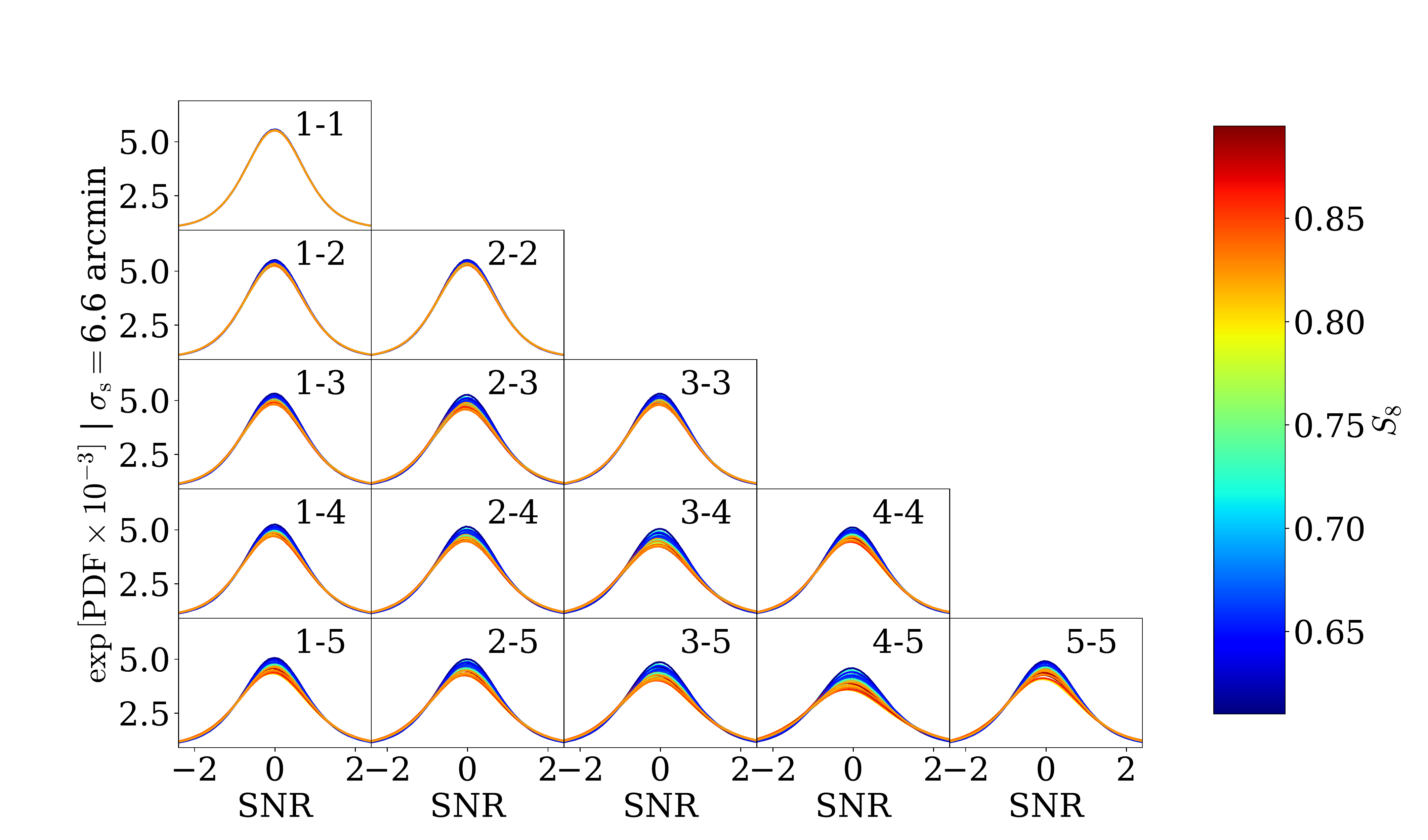} \\
\caption{The average lensing PDFs, as a function of signal-to-noise ratio (SNR), measured from the 26 cosmologies in the KiDS-1000 version of cosmo-SLICS, with smoothing scale $\sigma_{\rm s} = 6.6$ arcmin. The five tomographic redshift bins with edges $[0.1,0.3,0.5,0.7,0.9,1.2]$ are denoted by numbers 1-5 in increasing redshift order. The PDFs are colour-coded by the value of $S_8$ in each simulation. We show the exponential of the PDFs only to accentuate the differences between the cosmologies for illustration purposes.} \label{fig:pdfs_nbody_field} 
\end{figure*}

The auto- and cross-PDFs measured with smoothing scale $\sigma_{\rm s} = 6.6$ arcmin for the 26 distinct cosmo-SLICS cosmologies are presented in Figure \ref{fig:pdfs_nbody_field}. Colour-coding these by their $S_8=\sigma_8\sqrt{\Omega_{\rm m}/0.3}$ values, we see a similar trend to Figure \ref{fig:PDFs_HP}, where cosmologies with higher matter energy density and/or amplitude of the matter power spectrum give rise to broader PDFs. We also see a greater $S_8$ dependence in the PDFs for the higher redshift bins, consistent with 
these redshift ranges being richer sources of cosmological information. We note however that by eye the differences between the PDFs from different cosmologies is small - we plot the exponential of the PDFs here to accentuate the variation. This suggests that the shape of the lensing PDFs measured from a survey with number densities similar to KiDS-1000 are mostly dictated by the shape noise in the field.

The finely-binned PDFs shown in Figure \ref{fig:pdfs_nbody_field} are not used in the posterior probability evaluations due to the limitations in the length of the data vector set by the finite number of SLICS realisations (715 for the KiDS-1000 version of the mocks) available for estimating the covariance matrix. Consequently, following our approach with the Gaussian and lognormal fields, we use four SNR bins for each of the 15 redshift bin combinations in this test, resulting in data vectors which are 60 elements long per smoothing scale, and 180 elements long in the case of the concatenated data vector for the three smoothing scales combined. One could in principle use more bins to sample higher-order moments but this also comes with the risk of increasing the noise in the estimated covariance matrix. We leave further investigation of this for future work.

As for the range of the SNR bins, we follow the same approach as in Section \ref{subsec:method_gaussian}, selecting a symmetric binning scheme containing the peak and approximately 2$\sigma$ of the volume enclosed by the PDFs measured from the fiducial cosmology (see Fig.~\ref{fig:cosmo-SLICS}) which we adopt as the data vector in our tests. As with the previous tests (Sections \ref{subsec:method_gaussian}--\ref{subsec:method_lognormal}), we find that overzealous sampling of the PDF tails again introduces numerical instabilities to the inversion of the covariance matrix and noisy likelihood surfaces as a result. We continue to use different SNR binning for the PDFs from different smoothing scales, reflecting the varying widths of these statistics (see Fig.~\ref{fig:PDFs_HP}), but keep the binning constant across the redshift bin combinations and of course, all cosmologies.

In contrast to the Gaussian and lognormal maps which could be generated relatively quickly for arbitrary cosmologies on a 2D grid, the computational expense of numerical simulations means we have only 26 cosmologies in cosmo-SLICS which happen to sample a 4D parameter space $(\Omega_{\rm m}, S_8, h, w_0)$. In order to generalise the lensing PDFs measured at these 26 nodes to arbitrary values of $\Omega_{\rm m}, S_8, h$ and $w_0$, we use a Gaussian process regression (GPR) emulator to interpolate between the simulation cosmologies. GPR emulators have been trained on the cosmo-SLICS suite in numerous studies \citep{harnois-deraps/etal:2019,davies/etal:2020,davies/etal:2021, martinet/etal:2021, heydenreich/etal:2021} and found to be capable of predicting weak lensing statistics with percent-level accuracy across much of the parameter space depicted in Figure \ref{fig:cosmo-SLICS}. Following the approach taken in these previous analyses we adopt a Gaussian functional form for our emulator's covariance function and we train on the natural logarithm of the PDFs rather than the PDF itself, the former being a smoother function which improves emulation accuracy\footnote{We refer the interested reader to Appendix A of \citet{harnois-deraps/etal:2019} for more detailed discussion of GPR emulation with cosmo-SLICS.}. Finally, we train on the PDFs from each of the three smoothing scales and fifteen redshift bins separately, meaning that 45 different emulators with distinct sets of hyperparameters are used to produce our lensing PDF predictions. We find our emulators are accurate at the level of 1-5\% depending on the smoothing scale, redshift bin and cosmology (see Appendix \ref{app:emu_acc} for further details).

Once again we compare the power of the lensing PDF with the standard analysis involving two-point statistics. Whilst weak lensing practitioners have at their disposal analytical prescriptions for these statistics as a function of cosmology, the non-linear scales continue to require calibration with numerical simulations. One of the leading solutions to this problem, readily applicable with standard-use software packages such {\sc Nicaea} \citep{kilbinger/etal:2009} and {\sc CAMB} \citep{hojjati/etal:2011}, is the {\sc Halofit} approach of \citet{takahashi/etal:2012}. They model the non-linear scales with a function featuring free parameters fit to simulations varying in cosmology. However, as noted by \cite{harnois-deraps/etal:2019}, the cosmological parameter space covered by the \citet{takahashi/etal:2012} simulations is somewhat narrower than that of cosmo-SLICS (Fig.~\ref{fig:cosmo-SLICS}). Hence accuracy of the {\sc Halofit} approach cannot be guaranteed across the full 4D $(\Omega_{\rm m}, S_8, h, w_0)$ parameter space sampled by our lensing PDFs. This means that a fair comparison of two-point statistics with the lensing PDFs is more readily facilitated by also measuring the former from cosmo-SLICS and interpolating to arbitrary cosmologies with a GPR emulator. 

We therefore proceed to measure from the simulations the two-point shear correlation functions given by
\begin{equation} \label{eqn:xi+-}
\widehat{\xi_{\pm}^{ij}}(\theta) = \frac{\sum_{\rm{ab}} w_{\rm{a}}^i w_{\rm{b}}^j \left[ e_\text{t}^i (\boldsymbol{\theta}_{\rm{a}}) e_\text{t}^j (\boldsymbol{\theta}_{\rm{b}}) \, \pm \, e_\times^i (\boldsymbol{\theta}_{\rm{a}}) e_\times^j (\boldsymbol{\theta}_{\rm{b}})
\right]}{
\sum_{\rm{ab}} w_{\rm{a}}^i w_{\rm{b}}^j } \,,
\end{equation}
where the $e_{\text{t}}$ and $e_\times$ denote the tangential and cross ellipticity components (Eq.~\ref{eqn:e_obs}) measured relative to the vector connecting galaxy pairs at angular coordinates $\theta_{\rm{a}}$ and $\theta_{\rm{b}}$ separated on the sky by the angle $\theta$ \citep{bartelmann/schneider:2001}. The shear correlation functions quantify the correlation in galaxy shapes and orientations as a function of angular separation by summing over galaxy pairs within an interval $\Delta\theta$ centred on $\theta$ with weights $w_{\rm{a,b}}$ reflecting the accuracy with which the ellipticities are measured (in the case of our simulations, all weights take the value unity). These correlations can be estimated within a single redshift bin $(i=j)$ or across redshift bins $(i \neq j)$, and whilst we model their cosmological dependence numerically with cosmo-SLICS, for completeness we present the theoretical prescription given by
\begin{equation} \label{eqn:xi+-theory}
\xi_\pm^{ij}(\theta) = \frac{1}{2\pi}\int \text{d}\ell \, \ell \,C^{ij}(\ell) \, J_{0,4}(\ell \theta) \, , 
\end{equation}
where $J_{0,4}(\ell\theta)$ denotes the zeroth and fourth order Bessel functions, used for the $\xi_+$ and $\xi_-$ shear correlation function components respectively, and $C^{ij}(\ell)$ is the lensing power spectrum (Eq.~\ref{eqn:Cell}). 

The use of the shear correlation functions as our comparison two-point statistics in this test, as opposed to the lensing power spectrum, is another difference between the $N$-body analysis presented here and those performed on the Gaussian and lognormal fields in the previous sections. This choice was motivated by the \cite{asgari/etal:2021} cosmic shear analysis of KiDS-1000 which used $\widehat{\xi_{\pm}^{ij}}(\theta)$ statistics and others based on them. We do not expect our benchmark cosmological constraints to vary significantly had we instead measured the $C^{ij}(\ell)$s with an equivalent range of scales.

Using the publicly-available correlation function code {\sc TreeCorr} \citep{jarvis/etal:2004, jarvis:2015}, we measure the $\widehat{\xi_{\pm}^{ij}}(\theta)$ from the shear catalogues directly using the same nine logarithmically-spaced angular separation bins adopted by \cite{asgari/etal:2021}, between 0.7 and 300 arcminutes. In order to match the scales used in the lensing PDF and shear correlation function analyses as closely as possible, we use only the $\widehat{\xi_{\pm}^{ij}}(\theta \ge \rm{min}[\sigma_{\rm s}])$, where $\rm{min}[\sigma_{\rm s}]=2.2 \, \rm{arcmin}$ is the smallest smoothing scale used on our simulated convergence maps. Mirroring the lensing PDF analysis, we measure the five auto-correlations and ten cross-correlations facilitated by the five tomographic redshift bins, and we train GPR emulators on each redshift bin combination and on $\xi_+$ and $\xi_-$ separately. Specifically we train the emulators to predict $\theta \times \xi_\pm$, this being a smoother function with a narrower range of values than the $\xi_\pm$ statistics themselves. The correlation function data vector is 210 elements long (15 redshift bin combinations $\times$ seven $\theta$ bins $\times$ two correlation function components). We once again estimate the covariance of this statistic using the 715 SLICS realisations. Finally, we also evaluate the combined constraints of the lensing PDF (all smoothing scales) \textit{and} the correlation functions by concatenating these statistics in the likelihood (producing vectors of 390 elements long in total).

The grid-based likelihood sampling method used in the 2D $(\Omega_{\rm m}, \sigma_8)$ parameter space with the Gaussian and lognormal fields (Sects.~\ref{subsec:method_gaussian}-\ref{subsec:method_lognormal}) is inefficient for the 4D space $(\Omega_{\rm m}, S_8, h, w_0)$ covered by the cosmo-SLICS. Hence we use the {\sc emcee} MCMC software package \citep{foreman-mackey/etal:2013} to sample the 4D posterior probability distributions, applying uniform priors on $[\Omega_{\rm m}, S_8, h, w_0]$ with widths equivalent to the ranges sampled by the cosmo-SLICS (see Fig.~\ref{fig:cosmo-SLICS}). Since our simulations are tailored to match the properties of the KiDS-1000, we scale the estimated covariance matrices by the ratio of the areas of the SLICS lightcone (100 deg$^2$) and this data set (1000 deg$^2$) in order to simulate the KiDS-1000 constraining power.

For the ``data" in this mock cosmic shear analysis, we measure PDFs and $\xi_\pm$ from all available realisations of the fiducial cosmology in cosmo-SLICS. This suppresses sampling variance and produces data vectors which correspond to a much larger survey than KiDS-1000, thereby encouraging our measured cosmological constraints to centre on the truth. This is not the case with real astronomical data, where our constraints would scatter about the true cosmology in accordance with the sampling variance in the more limited KiDS-1000 sky coverage. Centring the constraints from the lensing PDFs and two-point statistics in this way, however, more readily facilitates a comparison of their relative constraining powers by eliminating the offsets in the constraints from the truth, which may vary in size and direction for PDFs versus $\xi_\pm$ with each possible 1000 deg$^2$ data realisation one could measure from the simulations.

\subsubsection{LSST analysis} \label{subsubsec:meth_lsst}

Whilst we are interested in the potential benefits of measuring the lensing PDF from current data sets, we also want to examine how any gains in cosmological precision will change with advancements in survey specifications. For this reason we repeat the analysis laid out in Section \ref{subsubsec:meth_k1000} but using versions of SLICS and cosmo-SLICS with galaxy number density almost five times higher than in the KiDS-1000 mocks, $n_{\rm gal}=30.5 \, \rm{arcmin^{-2}}$ $(z \in [0.1,1.2])$, similar to that which will be observed with the upcoming Vera Rubin Legacy Survey of Space and Time \citep{lsst/etal:2012}. Besides the overall galaxy density, the only other distinction between these versions of the SLICS and cosmo-SLICS is the redshift distribution of galaxies illustrated by Figure \ref{fig:nofz}. Statistically speaking, the deviations between the redshift distributions are relatively small, with the proportion of galaxies in each tomographic redshift bin in the KiDS-1000 and LSST simulation suites being within 3-8\% percent of each other, depending on the bin. Furthermore, we have no prior reason to suspect the differences in redshift distribution will notably affect the relative performance of the lensing PDFs and shear correlation functions in constraining cosmological parameters. We therefore attribute the differences we do observe between the relative constraining power of these probes primarily to the increase in signal-to-noise when changing from the KiDS-1000 to the LSST mocks.

  \begin{figure}
\centering
\includegraphics[width=0.49\textwidth]{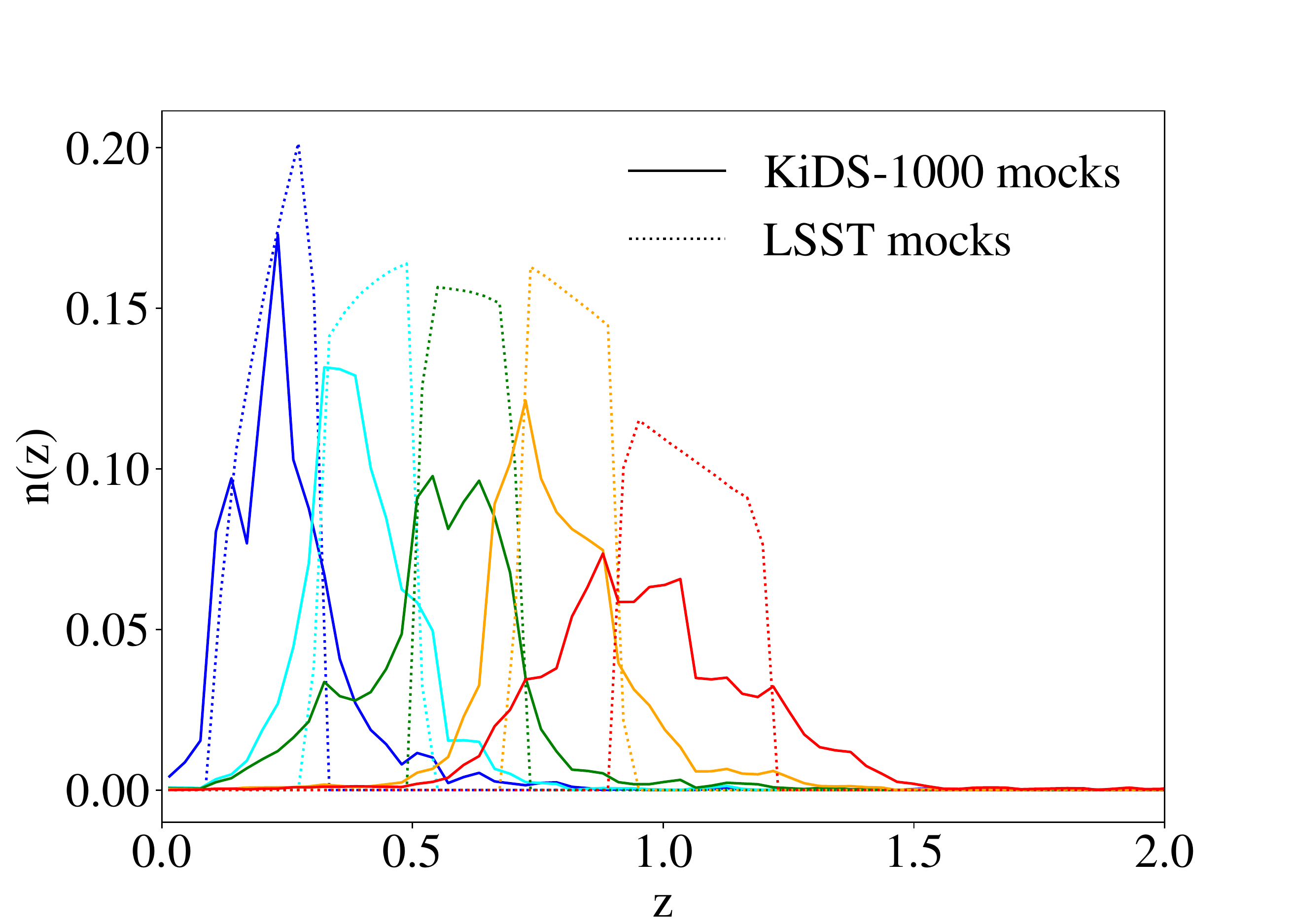} \\
\caption{Comparison of the normalised redshift distributions from the KiDS-1000 (solid lines) and LSST (dashed) versions of the simulations in each of the five tomographic bins (bordered by $[0.1,0.3,0.5,0.7,0.9,1.2]$; various colours). The tails in the KiDS-1000 mock $n(z)$s extending beyond the bounds of each tomographic bin arise from the fact that these are tailored to match those estimated for the most recent KiDS data \citep{hildebrandt/etal:2021}, where galaxies are binned based on their photometric redshift estimates with the distributions being in spectroscopic redshift.} \label{fig:nofz} 
\end{figure}

The upcoming LSST will of course exceed Stage-III lensing surveys not only in observed galaxy number density but also in sky coverage. In order to simulate this additional enhancement in survey specifications in our analysis, we scale the covariance matrices in this test to correspond to an area of 18,000 deg$^2$, approximately that which will be surveyed by LSST. This means that our comparison of the PDF-to-$\xi_\pm$ cosmological constraint improvements between the KiDS-1000 and LSST analyses is a test of two effects simultaneously - an increase in galaxy number density and in survey area. This allows for a more realistic forecast of how effective the lensing PDF will be in extracting additional cosmological information over standard two-point statistics, from future surveys relative to those of the present day.

One final deviation between the KiDS-1000 and LSST mocks is that we have slightly fewer (616) realisations of the latter available for computing the covariance matrices (previously 715). This is still larger than the length of the longest data vector (390) appearing in our likelihoods, and is therefore sufficient for covariance estimation. As in Sects.~\ref{subsec:method_gaussian}-\ref{subsec:method_lognormal}, we apply the appropriate \cite{hartlap/etal:2007} scaling factors to the covariance matrices to mitigate any potential biases occurring in the estimation from our finite number of realisations. As these factors scale with the number of simulations, the factors vary slightly between the KiDS-1000 and LSST analyses.

\section{Results} \label{sec:results}

\subsection{KiDS-1000} \label{subsec:results_k1000}

  \begin{figure*}
\centering
\includegraphics[width=0.63\textwidth]{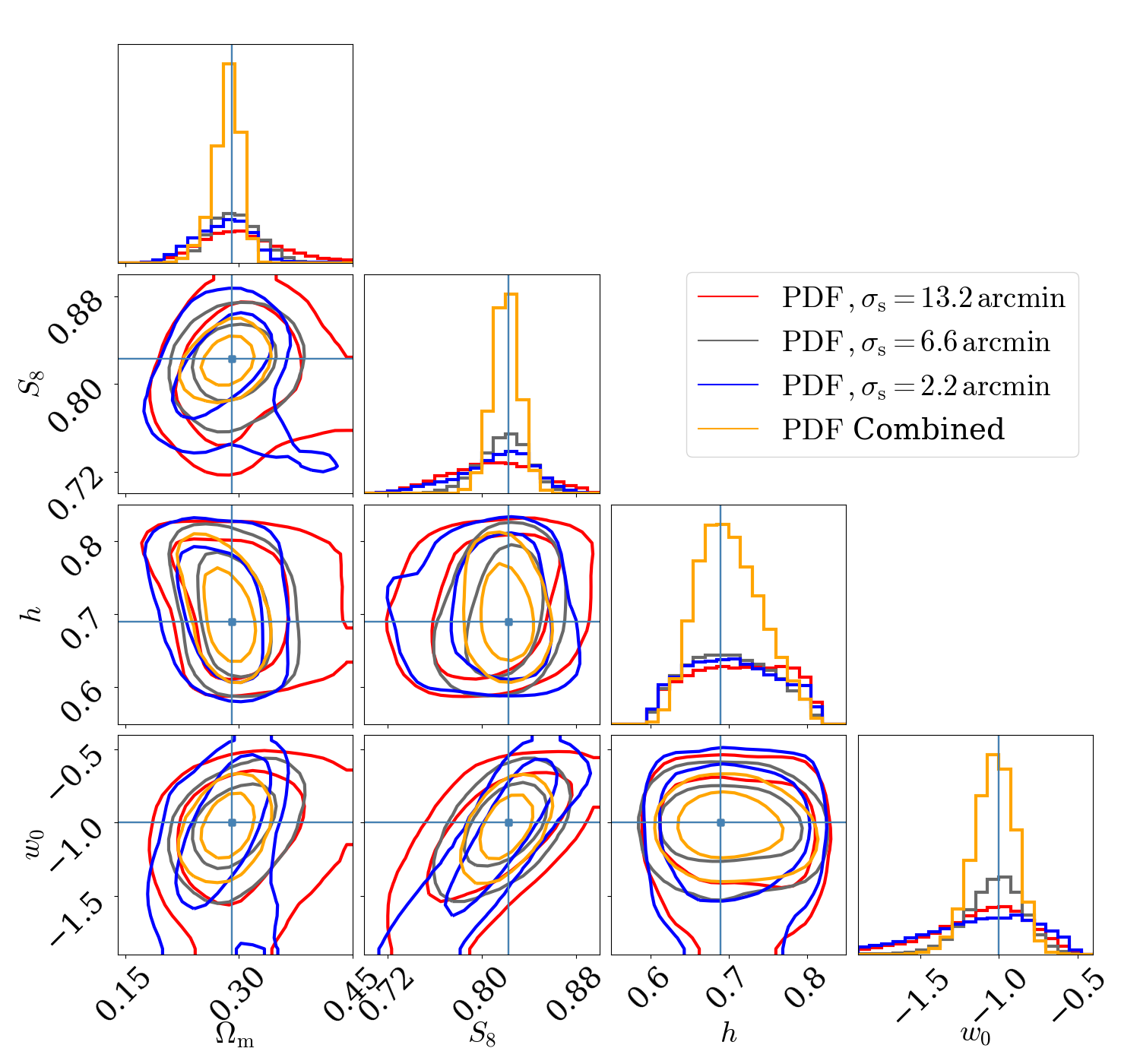} \\
\includegraphics[width=0.63\textwidth]{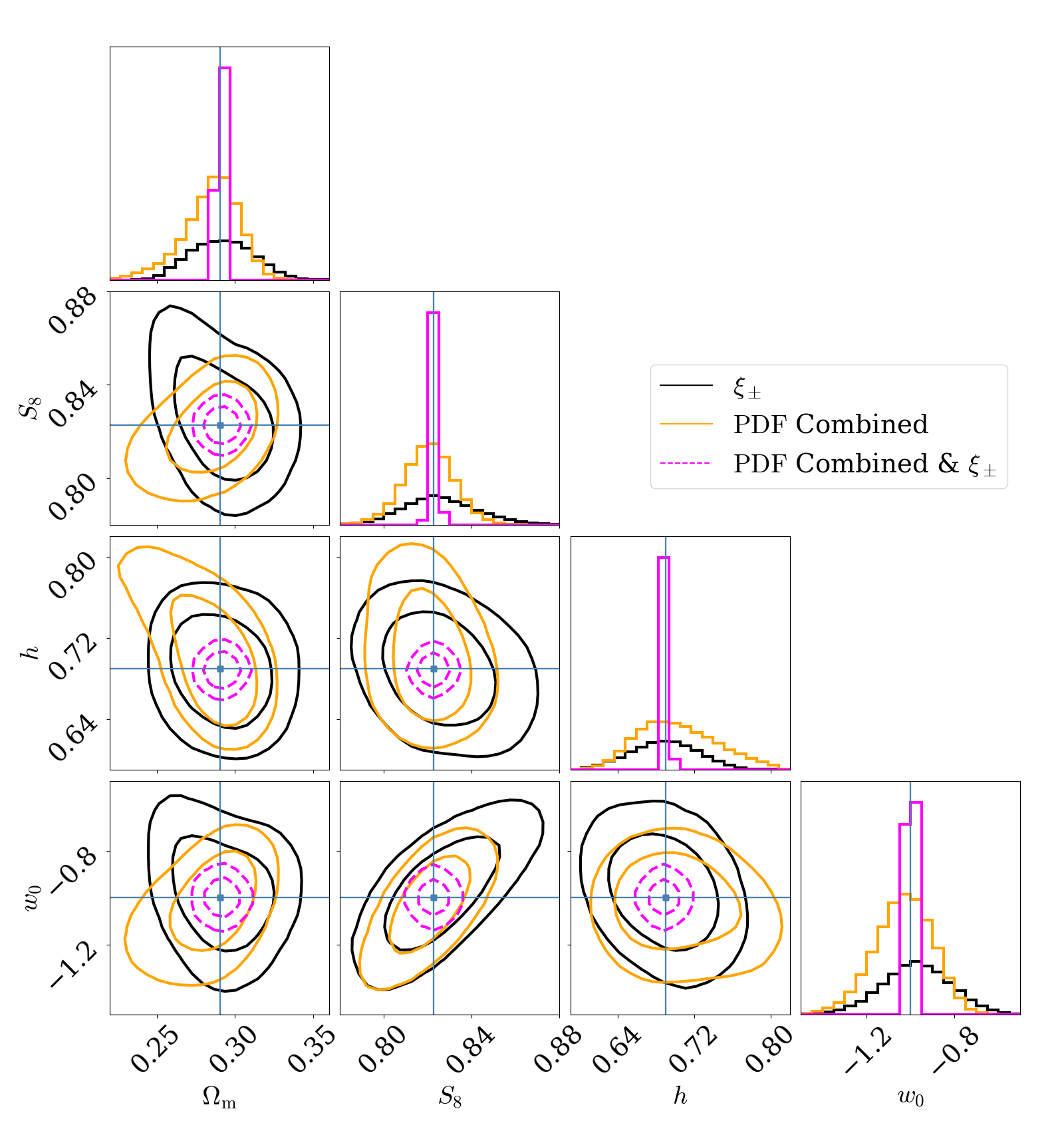} \\
\caption{Cosmological parameter constraints measured from the KiDS-1000 simulated survey (galaxy number density $n_{\rm gal}=6.2 \, {\rm arcmin}^{-2}$ and survey area $A_{\rm survey} = 1000 \, {\rm deg}^2$). \textit{Upper:} The individual constraints from the lensing PDFs at the three smoothing scales (red, grey, and blue lines), and the combined constraints from all three (orange). \textit{Lower:} the same combined-PDF constraints compared with the two-point shear correlation function constraints (black), and the combined constraints of all the statistics (dashed magenta). Note that the axes limits are narrower in the lower panel to better illustrate the extent of the contours. } \label{fig:constraints_nbody_field} 
\end{figure*}

Figure \ref{fig:constraints_nbody_field} presents the cosmological constraints derived from our mock KiDS-1000 analysis\footnote{The corner plots presented in Figures \ref{fig:constraints_nbody_field}  and \ref{fig:constraints_K1000vsLSST} were produced with the corner.py software from \citet{foreman-mackey:2016}.}. The upper panel displays those measured from the lensing PDFs, with the three individual smoothing scale statistics in blue ($\sigma_{\rm s} = 2.2 \, \rm{arcmin}$), grey ($6.6 \, \rm{arcmin}$), and red ($13.2 \, \rm{arcmin}$) respectively, and their combination in orange. The lower panel compares the combined PDF constraint with those of the two-point shear correlation functions (in black) and with the combined constraint of all lensing PDFs \textit{and} the $\xi_\pm$ (dashed magenta). Henceforth, when presenting numerical comparisons of the relative sizes of the constraints from the various probes, we do so using the 68\% confidence intervals.

From the increased precision offered by the combined PDF constraints relative to those from the individual smoothing levels, we can immediately infer that there is independent information contained in the $\sigma_{\rm s} = 2.2, \, 6.6$ and 13.2 arcmin measurements, consistent with the findings of \citet{boyle/etal:2021} \new{and \citep{kratochvil/etal:2012}}. Whilst we only measured this effect for the $\Omega_{\rm m}$ and $\sigma_8$ parameters in the Gaussian and lognormal test cases (Sects.~\ref{subsec:method_gaussian}-\ref{subsec:method_lognormal}), here we observe this also for the Hubble and dark energy equation of state parameters. In terms of the balance of information between the three smoothing scales, the biggest contribution comes from the PDF with the intermediate level of smoothing, 6.6 arcmin. Whilst the lower smoothing scale, 2.2 arcmin, allows the lensing PDF to probe smaller physical scales, the increase in noise we observe in the less-smooth convergence maps (see Fig.~\ref{fig:kappa_maps}) overwhelms any potential improvement in the constraining power from the extra small-scale signal. This is evident from the noisy contours we obtain for this level of smoothing. Likewise, the largest smoothing scale, 13.2 arcmin, suppresses spurious noise features but at the expense of the cosmological information contained in the smaller scales. Hence we find that 6.6 arcmin provides the best compromise of these effects, yielding constraints on $S_8$ which are 35\% and 46\% tighter than the lower and higher smoothing scales respectively.

Nevertheless, we find that including the other two smoothing scale measurements in the likelihood is a notable improvement on the 6.6 arcmin PDF alone; the combined PDF yields constraints on $\Omega_{\rm m}$, $S_8$, $h$ and $w_0$ which are 25-48\% tighter, depending on the parameter, compared to those from the intermediate smoothing scale. Furthermore we find that all smoothing scales contribute to the gain in precision, such that the combination all three yields tighter constraints than any pair of smoothing scales.

From the lower panel of Figure \ref{fig:constraints_nbody_field}, and the metrics presented in the upper-middle column of Table \ref{tab:constraints_k1000_lsst}, we see that the combined PDF offers tighter constraints on $\Omega_{\rm m}, S_8$, and  $w_0$ than the shear correlation functions, by 18\%, 33\%, and 16\% respectively. However, the conventional probe outperforms the lensing PDF in constraining the Hubble parameter by 17\%. \new{
The discrepancy on $h$ could be influenced by the prior rather than a difference in the constraining power of the two statistics. Indeed, we see that the extent of the PDF and $\xi_\pm$  constraints approaches the prior boundaries on $h$ (given by the range of the nodes shown in Fig.~\ref{fig:cosmo-SLICS}), and cosmic shear generally places very weak constraints on the Hubble parameter \citep{hall:2021}. Whilst this effect could be tested in principle by widening the prior bounds, in this case this would lead to emulating predictions outside of the cosmo-SLICS parameter space where the model's accuracy is unverifiable, potentially leading to unreliable constraints. We therefore advise a note of caution in interpreting this difference in the $h$ constraints, and place greater emphasis on the higher signal-to-noise results obtained from the combination of the lensing PDFs and $\xi_\pm$, and with the LSST-like survey in Section \ref{subsec:results_lsst}.
}

The overall combination of the lensing PDFs with the $\xi_\pm$ presented in the lower panel of Figure \ref{fig:constraints_nbody_field} (dashed magenta), reveals that these two probes indeed contain independent information. Using both the $\xi_\pm$ and lensing PDF in the likelihood results in improvements in the $\Omega_{\rm m}, S_8, h$, and $w_0$ constraints by 30\%, 26\%, 38\% and 33\% respectively compared to the lensing PDF alone. When we compare the overall combined constraints to those from the $\xi_\pm$ alone, the precision gains are even more encouraging: 43\%, 51\%, 28\% and 44\% tighter constraints on these four parameters are obtained. These metrics are also presented in the lower-middle column of Table \ref{tab:constraints_k1000_lsst} for ease of reference.

\new{
The improvement in cosmological constraints in the PDF-$\xi_\pm$ combined analysis arises here in part from differing degeneracy directions in the statistics individually. For example, the relative orientations of the orange combined-PDF and black $\xi_\pm$ contours in the $\Omega_{\rm m}-w_0$ plane is qualitatively similar to those shown in the (closely-related) $\Omega_{\rm cdm}$ (cold dark matter)$-w_0$ plane in Fig. 11 of \cite{boyle/etal:2021}. We also ran MCMCs sampling $\sigma_8$ instead of $S_8$, finding the familiar ``banana-shaped" $\Omega_{\rm m}-\sigma_8$ degeneracy is reproduced with both statistics, consistent with \cite{boyle/etal:2021}. In these tests the relative constraining power of the lensing PDF and $\xi_\pm$ are practically unchanged from the $S_8$-sampling chains.
}

It is interesting to note that in the case of this simulated KiDS-1000 survey, the combination of PDFs (all scales) and two-point statistics proves to be more constraining than the combined PDF only, whereas in Section \ref{subsec:method_lognormal} we saw with the lognormal field that the two-point statistic contributed no additional constraining power to the PDF analysis. The reason for this difference, as we shall see in the following section with the LSST-like mock analysis, comes down to the level of noise in the field. In the high signal-to-noise regime, as was the case with the noise-free lognormal maps, the lensing PDF performs significantly better than in a Stage-III lensing survey scenario, where the galaxy shape noise and limited galaxy number density swamp the cosmological information available to the PDF. We revisit this point in Section \ref{subsec:results_lsst}.

\begin{table}
  \begin{center}
    \caption{Changes in the size of the 68\% confidence intervals offered by the lensing PDF relative to the $\xi_\pm$ (upper half of the table), and by the combination of the lensing PDF and $\xi_\pm$ relative to $\xi_\pm$ alone (lower half), measured from the KiDS-1000 and LSST simulated surveys. Positive numbers indicate tighter constraints than $\xi_\pm$ alone, and vice versa. } \label{tab:constraints_k1000_lsst}
  \begin{tabular}{|lcr|}
\hline
\hline
Parameter & KiDS-1000  & LSST \\
$\Omega_{\rm m}$ & 18\% & 85\% \\
$S_8$            & 33\% & 90\% \\
$h$              & -17\% & 93\% \\
$w_0$            & 16\% & 97\% \\
\hline
$\Omega_{\rm m}$ & 43\% & 86\% \\
$S_8$            & 51\% & 91\% \\
$h$              & 28\% & 94\% \\
$w_0$            & 44\% & 98\% \\
\hline \hline
  \end{tabular}
  \end{center}
\end{table}

\subsection{LSST} \label{subsec:results_lsst}

  \begin{figure*}
\centering
\includegraphics[width=0.68\textwidth]{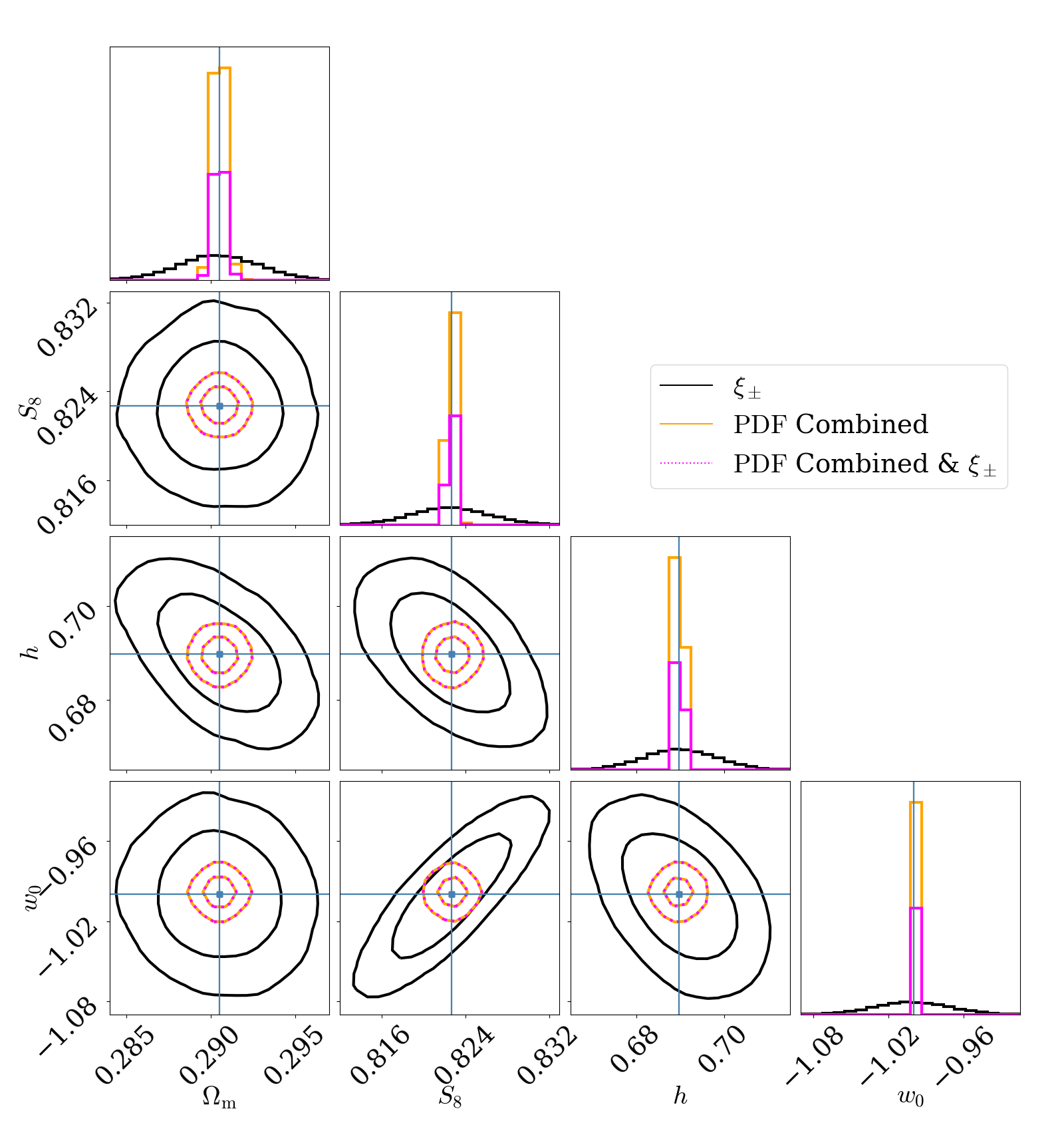} \\
\caption{The same as the lower panel of Fig.~\ref{fig:constraints_nbody_field} but measured from the LSST simulated survey ($n_{\rm gal}=30.5 \, {\rm arcmin}^{-2}$, $A_{\rm survey} = 18,000 \, {\rm deg}^2$). Note that the orange and dashed magenta contours are almost perfectly superimposed. } \label{fig:constraints_K1000vsLSST} 
\end{figure*}

With the improvements in constraining power offered by the lensing PDF established for a survey similar to KiDS-1000, we now proceed to examine how these gains might change with future survey developments. Figure \ref{fig:constraints_K1000vsLSST} presents the constraints from the $\xi_\pm$ (in black), lensing PDF (all smoothing scales; orange), and the overall combination (dashed magenta, almost exactly on top of the orange contours), for an LSST-like survey. These are measured from our simulations with galaxy number density $30.5 \, \rm{arcmin}^2$ and with the covariance matrices scaled to correspond to a survey area of 18,000 deg$^2$.

First of all, we note that the contours which use the lensing PDFs in this test appear to be quite circular and do not display strong parameter degeneracies. This could be partially down to the extra cosmological information extracted by the lensing PDFs over the two-point statistics, but could also be caused by the limited parameter sampling in the cosmo-SLICS (see Fig.~\ref{fig:cosmo-SLICS}). As a result of the increase in galaxy number density and survey area in this test, only the fiducial cosmology node is contained within the 95\% confidence intervals produced with the lensing PDFs. This means that the emulators' predictions in this part of the parameter space are derived strongly from the fiducial cosmology simulation measurement. This could in principle mean that the shape and area of the observed posterior may differ from the true posterior we would obtain if we had an arbitrarily-dense array of nodes from which summary statistics could be measured. However, this effect is likely to inflate rather than reduce the size of the contours. Since the $\xi_\pm$ constraints are larger and more strongly sample the other training nodes, it is less affected by this issue. Therefore the comparison of the PDF-derived constraints to those of the $\xi_\pm$ will produce a conservative estimate of the relative constraining power of the lensing PDFs. Having more nodes to sample the cosmological parameters in the future will shed light on this.

Secondly, we once again find that the combined lensing PDF offers a smaller contour area than the $\xi_\pm$, but the benefits of the lensing PDF have appreciated significantly for all cosmological parameters with the increase in galaxy number density and survey area. This is apparent from the relative size of the lensing PDFs and $\xi_\pm$ contours in this figure compared to those in Fig.~\ref{fig:constraints_nbody_field}, and from the quantitative comparison presented in the right column of Table \ref{tab:constraints_k1000_lsst}. The improvements in the $\Omega_{\rm m}$, $S_8$, and $w_0$ constraints offered by the lensing PDFs relative to $\xi_\pm$ in the KiDS-1000 simulated survey, 18\%, 33\% and 16\% respectively,  have all significantly improved, to 85\%, 90\% and 97\% respectively. Even the Hubble parameter, which was 17\% more precisely constrained by $\xi_\pm$ in the KiDS-1000 test, is 93\% more tightly constrained by the PDFs than the standard two-point correlation functions in the LSST survey.

Figure \ref{fig:constraints_K1000vsLSST} also shows that the combination of the lensing PDFs and $\xi_\pm$ statistics produces constraints which are practically superimposed with those of the lensing PDFs alone. This means that the $\xi_\pm$ adds almost zero extra information to the analysis and that all of the constraining power comes from the PDFs. This can also be seen in Table \ref{tab:constraints_k1000_lsst}, where the metrics pertaining to the combination of the PDFs and $\xi_\pm$ (lower half of the table) are almost unchanged from those pertaining to the PDFs alone (upper half). This result differs notably from our findings with the simulated KiDS-1000 analysis presented in the previous section, where the addition of the $\xi_\pm$ improved the precision of lensing PDF cosmological constraints by tens of percent, but strongly echoes our observations in Section \ref{subsec:method_lognormal} with the noise-free lognormal field. The common denominator in these cases is the signal-to-noise; when it is high, the two-point statistics contain only a subset of the information harnessed by the lensing PDFs.

The reason that the lensing PDFs improves  in constraining power relative to the two-point statistics as the signal-to-noise of the survey is increased, is down to the sensitivity of the former measurement to the galaxy shape noise, discussed in Section \ref{subsubsec:meth_k1000}. With a KiDS-1000 galaxy density and shape noise level, we find that the lensing PDFs are quite noise dominated, especially in the lower redshift bins as shown in Figure \ref{fig:pdfs_nbody_field}. With the increased galaxy density however, the noise dispersion in the convergence maps decreases (see Eq.~\ref{eqn:kappa_noise}) leading to more signal-dominated maps and greater differences in the lensing PDFs measured from the various cosmo-SLICS cosmologies. The emulator is subsequently better positioned to learn the cosmological dependence of these new statistics and distinguish different parameter configurations in the course of the parameter inference via MCMC. The shape of the two-point shear correlation functions on the other hand is less sensitive to the signal-to-noise of the field, meaning that one gains less cosmological precision from this statistic  - compared to the lensing PDF  - as the survey specifications are improved.


\section{Conclusions} \label{sec:conclusions}

We have investigated the cosmological constraining power of an alternative cosmic shear statistic, the lensing PDF, consisting of histograms of the projected matter density smoothed with filters of various widths. We have established intuition on the cosmological information contained in this new statistic step by step, beginning with simple Gaussian and lognormal fields before constructing an accurate numerical model of the tomographic lensing PDF based on high-fidelity simulations. This has allowed us to probe smaller scales than ever before with the lensing PDF, unlocking new potential for constraining cosmological parameters. 

With the Gaussian density fields we verified key results from \citet{boyle/etal:2021} \new{and \citep{kratochvil/etal:2012}}, that the lensing PDFs measured from maps smoothed with filters of different widths contain different and complimentary cosmological information. By combining the PDFs from a small number of smoothing scales, the cosmological precision approaches that of the two-point lensing power spectrum, the optimal statistic for constraining the cosmology of a Gaussian field. With lognormal density fields however, we find that the combined lensing PDF considerably outperforms the power spectrum, confirming that the former statistic accesses the non-Gaussian information which is beyond the reach of the conventional two-point probe. 

We then demonstrated the benefits of the lensing PDFs in more realistic scenarios, using numerical weak lensing simulations with numerous input cosmologies and specifications tailored to match current and future weak lensing surveys - SLICS \citep{harnois-deraps/etal:2018} and cosmo-SLICS \citep{harnois-deraps/etal:2019}. We trained a Gaussian process emulator to learn the cosmological dependence of the lensing PDFs from cosmo-SLICS and subsequently performed a mock systematics-free KiDS-1000 cosmic shear analysis including galaxy shape noise, tomographic redshift binning, and angular scales as small as 2.2 arcminutes. We found the lensing PDF approach yields constraints on $\Omega_{\rm m}$, $S_8$ and $w_0$ which are 16-33\% tighter, depending on the parameter, than the conventionally-used shear correlation functions, $\xi_\pm$, when the angular scales used in each statistic are matched as closely as possible. The correlation functions continued to constrain  the Hubble parameter more precisely however, by 17\%. On the other hand, the combination of the lensing PDFs (all smoothing scales) \textit{and} the $\xi_\pm$ led to a 51\% gain in precision for the $S_8$ parameter over $\xi_\pm$ alone, with benefits of 28-43\% found for the other three cosmological parameters.

Finally we investigated how the relative performance of the lensing PDFs and $\xi_\pm$ will change as weak lensing surveys advance. We simulated a systematics-free LSST-like cosmic shear analysis by measuring the lensing PDFs and correlation functions from versions of the SLICS and cosmo-SLICS with galaxy number density approximately five times higher than in the version of the mocks tailored to match KiDS-1000, and scaling the covariance matrices to correspond to an 18,000 deg$^2$ survey. We showed that the lensing PDFs benefit more strongly from the increase in signal-to-noise than the $\xi_\pm$ on account of the sensitivity of the former statistic to the ratio of the galaxy shape noise dispersion and the number density. This means that the benefits of the lensing PDFs over $\xi_\pm$ appreciated considerably, ranging from 85\% (for $\Omega_{\rm m}$) to 97\% (for $w_0$), with the $S_8$ constraints weighing in at 90\% tighter than those from $\xi_\pm$. We therefore expect the advantages of this novel statistic to grow with the upcoming survey developments, increasing the impetus for adopting the lensing PDF as a conventional cosmic shear probe.

Furthermore, contrary to our findings with the KiDS-like simulated survey, in the LSST case the lensing PDF constraints remain practically unchanged when this statistic is combined with the $\xi_\pm$. This result re-enforces our findings with the noise-free lognormal test case: namely, with higher signal-to-noise surveys the lensing PDFs capture all of the cosmological information contained in the standard two-point probe as well as previously-inaccessible non-Gaussian information. 

Whilst the lensing PDF is clearly a promising tool for improving cosmic shear constraints, its sensitivity to the numerous sources of weak lensing systematics is currently not fully understood. The first steps in understanding the impact of systematics including intrinsic alignments, baryonic feedback, and survey footprint geometry on weak lensing peak counts were taken in a number of recent works \citep{kacprzak/etal:2016, martinet/etal:2018, shan/etal:2018, harnois-deraps/etal:2021}. They present effective frameworks for modelling these effects on the lensing PDF which could be implemented in the future. The question of systematics and the biases they potentially manifest hangs over all weak lensing probes, and methods to effectively model and marginalise over these effects remains a challenging pursuit. Given the compelling gains in cosmological precision facilitated by the lensing PDF and the evidence for the appreciation of these gains with future surveys, tackling the impact of systematics on this statistic is clearly a worthwhile endeavour. 

Finally, for the future benefits of the lensing PDF to be realised in practice for surveys such as LSST, increasingly accurate models for its cosmological dependence will be required. The analytical approaches of \citet{thiele/etal:2020} and \cite{boyle/etal:2021} present computationally inexpensive means to do this but at the expense of missing out on the information contained on the smaller scales. Our emulator approach using numerical simulations renders the smaller scales accessible, but larger simulation suites with more nodes will be required to model the lensing PDF at the percent-level accuracy required for upcoming Stage-IV surveys. The CosmoGrid suite \citep{kacprzak/etal:2022}, for example, comprises simulations with 2500 unique cosmologies, but they are currently designed for analysis of Stage-III photometric surveys and lack the small-scale resolution offered by the cosmo-SLICS. Nevertheless, suites such as these provide an excellent foundation on which to base further simulation development going forward.

\section*{Acknowledgements}

BG acknowledges the support of the Royal Society through an Enhancement Award (RGF/EA/181006) and the Royal Society of Edinburgh for support through the Saltire Early Career Fellowship (ref. number 1914). 
YC acknowledges the support of the Royal Society through a University Research Fellowship and an Enhancement Award. YC thanks the hospitality of the Astrophysics and Theoretical Physics groups of the Department of Physics at the Norwegian University of Science and Technology during his visit. 
JHD is supported by an STFC Ernest Rutherford
Fellowship (project reference ST/S004858/1). This work was made possible thanks to the HPC facilities supported by Eric Tittley at the IfA, Edinburgh. 
The authors are grateful to Ludo van Waerbeke for the mass reconstruction code used in this work, and to the anonymous referee for their valuable and helpful input on the first manuscript. 

\section*{Data Availability}

The SLICS lensing simulation suite is freely available at \url{https://slics.roe.ac.uk/}, and the cosmo-SLICS are available upon request. The Gaussian process regression emulator code used in this work is accessible at \url{https://github.com/benjamingiblin/GPR_Emulator}.

\bibliographystyle{mnras}
\bibliography{references}

\medskip

\appendix
\section{Emulator accuracy} \label{app:emu_acc}

We evaluate the accuracy of our emulators' lensing PDF predictions for each redshift bin combination and smoothing scale using leave-one-out cross-validation (CV). This consists of cycling through the 26 cosmo-SLICS cosmologies, omitting each one from the training set and re-training on the remaining 25 before making a prediction for the missing cosmology. By comparing the prediction to the simulated measurement from the omitted cosmology we gain an estimate of the accuracy of the emulators used in our MCMCs. This estimate is a conservative one, given that the MCMC emulators have the advantage of being trained on all available cosmologies. We define accuracy as the fractional difference between the prediction and the simulated measurement. Hence, positive and negative accuracies correspond to over- and underestimates by the emulator respectively, with null accuracies corresponding to perfectly unbiased predictions. 

  \begin{figure*}
\centering
\includegraphics[width=0.99\textwidth]{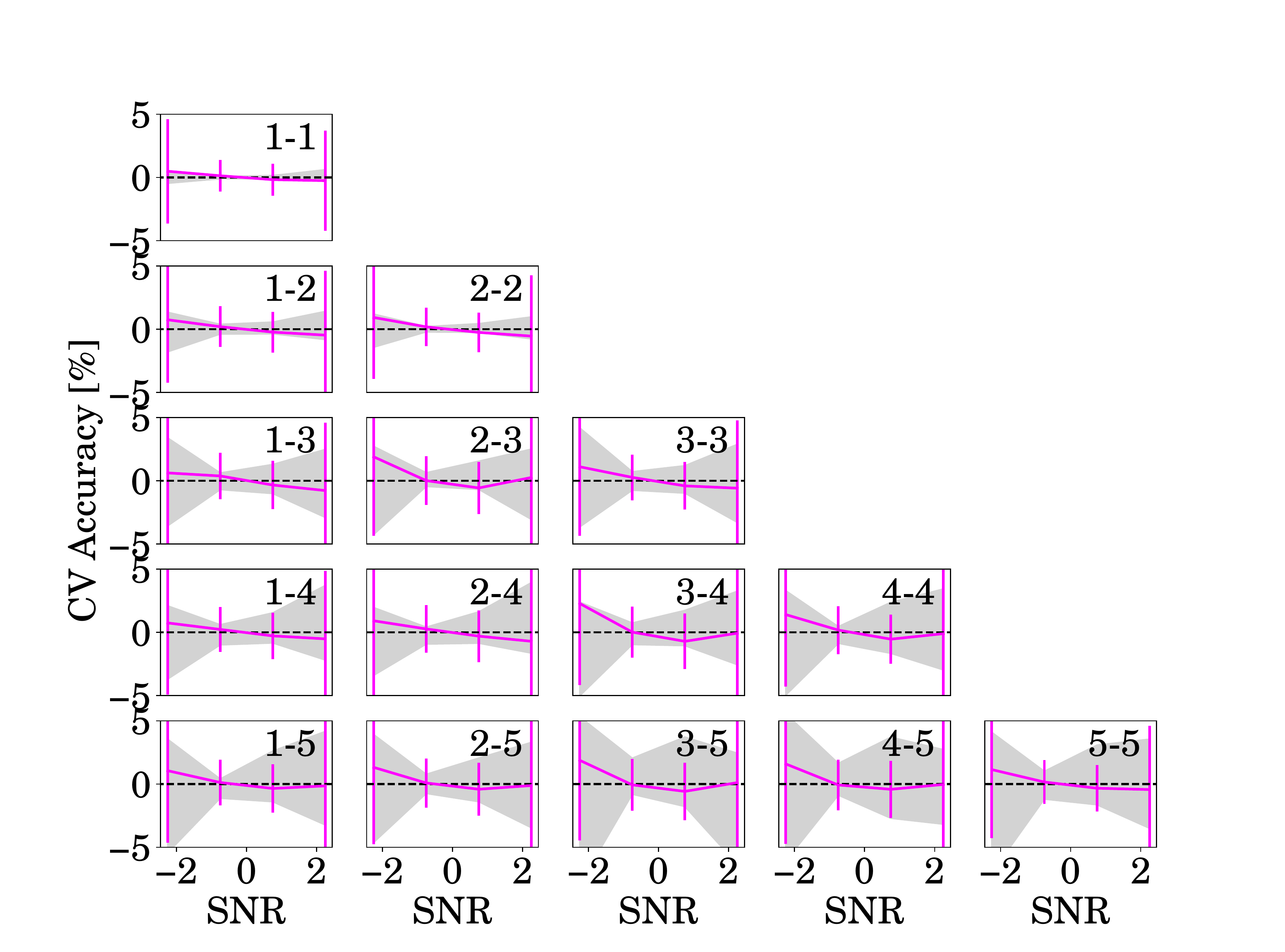} \\
\caption{The accuracy of the KiDS-1000-like PDF $(\sigma_{\rm s}=6.6 \, \rm{arcmin})$ emulator predictions in each redshift bin evaluated by cross-validation. The dashed line at zero indicates perfect accuracy, with positive and negative accuracies corresponding to over- and underestimates by the emulator respectively. The magenta line shows the result for the fiducial cosmology - used as the data vector in our MCMCs - with the error bars indicating the standard deviation measured from SLICS rescaled to correspond to a 1000 deg$^2$ survey. The grey region shows the range of accuracies obtained for the other 25 cosmo-SLICS cosmologies. The numerical labelling of the tomographic bins follows Fig.~\ref{fig:pdfs_nbody_field}.} \label{fig:cv_acc} 
\end{figure*}

Figure \ref{fig:cv_acc} shows the CV accuracies obtained for the 6.6 arcmin lensing PDFs measured from the KiDS-1000 version of cosmo-SLICS. The magenta line denotes the fiducial cosmology used as the data vector in our MCMCs, with the error bars indicating the statistical uncertainty attributed to the data in our parameter inference. This consists of the standard deviation measured with the 715 SLICS realisations, rescaled to correspond to the 1000 deg$^2$ survey area of KiDS-1000. The grey bands show the range of accuracies obtained for the other 25 cosmo-SLICS cosmologies.

We can see that the accuracies are generally better in the lower redshift bins. This is because the PDF measurements from these redshifts are dominated by shape noise and there is scarcely any cosmological variation for the emulators to learn and replicate. Secondly, we note that the deviation between the predicted and simulated measurements for the fiducial cosmology, shown by the difference between the magenta and zero lines, is consistently smaller than the error bars. This means that the biases in the emulators' predictions are subdominant to the statistical noise present in a KiDS-1000-like survey. 

Finally we find that there is generally a small spread in the accuracies for the other cosmologies, shown by the grey band, which is centred on the zero line denoting perfect predictions.  We expect to measure greater biases for the cosmologies at the edge of the cosmo-SLICS parameter space (see Fig.~\ref{fig:cosmo-SLICS}) since these are not completely enclosed by neighbouring nodes. Nevertheless, the predictions for all cosmologies are generally accurate at the level of a few percent. 

We have also checked the accuracies for the other smoothing scales measured from the KiDS-1000 mocks, as well as those from the LSST version of cosmo-SLICS. In the latter case we compare the size of the emulator biases with the statistical uncertainty used in the cosmic shear analysis of the simulated LSST data (the covariance scaled to represent an 18,000 deg$^2$ survey; see Sect.~\ref{subsubsec:meth_lsst}). In all cases we obtain qualitatively similar results, with 2.2 arcmin and 13.2 arcmin smoothing yielding accuracies which are a few percent better and a few percent worse, respectively, than those of the intermediate smoothing scale. The fact that the emulators in all cases have relatively very small biases for the cosmology used as the data in our MCMCs, and that the majority of predictions are accurate to better than 5\%, are strong indicators that the emulators trained on cosmo-SLICS and employed in this work are sufficiently accurate to use in cosmic shear analyses. These findings are consistent with other studies which employ these simulations to train GPR emulators \citep{harnois-deraps/etal:2019,davies/etal:2020,davies/etal:2021,heydenreich/etal:2021}.

\end{document}